\documentclass[usegraphicx,useAMS,usenatbib,fleqn]{mn2e}
\include{latex_macros}
\usepackage{float}

\usepackage{times}
\usepackage{amsmath}
\usepackage{amssymb}
\usepackage{amsfonts}
\usepackage{mathrsfs}
\title {The inflated radii of M-dwarfs in the Pleiades}
\author[R. J. Jackson, C. P.~Deliyannis and R. D. Jeffries]
  {R. J.~Jackson$^1$, Constantine P.~Deliyannis$^2$ and R. D.~Jeffries$^1$\\
   $^1$ Astrophysics Group, Keele University, Keele, 
      Staffordshire ST5 5BG\\
     $^2$Department of Astronomy, Indiana University.
727 E 3rd Street,
Bloomington, IN 47405-7105, USA
}
\date{In press}

\pagerange{\pageref{firstpage}--\pageref{lastpage}} \pubyear{2018}

\def\LaTeX{L\kern-.36em\raise.3ex\hbox{a}\kern-.15em
    T\kern-.1667em\lower.7ex\hbox{E}\kern-.125emX}

\begin{document}
\label{firstpage}
\maketitle

\begin{abstract}
Rotation periods obtained with the Kepler satellite have been combined
with precise measurements of projected rotation velocity from the WIYN 3.5-m
telescope to determine the distribution of projected radii for several
hundred low-mass ($0.1 \leq M/M_{\odot} \leq 0.8$), fast-rotating
members of the Pleiades cluster. A maximum likelihood modelling
technique, that takes account of observational uncertainties, selection
effects and censored data, and considers the effects of differential rotation and
unresolved binarity, has been used to find that the average radius of
these stars is $14\pm2$ per cent larger at a given
luminosity than predicted by the evolutionary models of Dotter et
al. (2008) and Baraffe et al. (2015). The same models are a reasonable
match to the interferometric radii of older, magnetically inactive
field M-dwarfs, suggesting that the over-radius may be
associated with the young, magnetically active nature of the Pleiades
objects. No evidence is found for any
change in this over-radius above and below the boundary marking the
transition to full convection.  Published evolutionary models that
incorporate either the effects of magnetic inhibition of convection or
the blocking of flux by dark starspots do not individually explain the
radius inflation, but a combination of the two effects might. The
distribution of projected radii is consistent with the adopted
hypothesis of a random spatial orientation of spin axes; strong
alignments of the spin vectors into cones with an opening semi-angle
$<30^{\circ}$ can be ruled out. Any plausible but weaker alignment
would increase the inferred over-radius.
\end{abstract}


\section{Introduction}
The radii of tidally-locked, main-sequence K- and M-dwarfs in
eclipsing binary systems are consistently measured to be larger than
predicted by most evolutionary models. The radius discrepancies amount to
10--20 per cent at a fixed mass and thus for a fixed luminosity,
the effective temperature, $T_{\rm eff}$, can be underestimated by 5--10
per cent (e.g. Lopez-Morales \& Ribas 2005; Morales et al. 2009; Torres
2013). The tidally-locked 
components of these binary systems are fast-rotating and magnetically
active; this, together with the fact that interferometrically
measured radii for nearby, relatively inactive K- and M-dwarfs are in
much better agreement with models (e.g. Demory et al. 2009; Boyajian et
al. 2012), has led to theoretical developments that explain ``radius
inflation'' in terms of the effects of dynamo-generated magnetic
activity (Morales, Ribas \& Jordi 2008). 

Magnetic activity might have an influence on the radii of cool,
convective stars either through inhibition of convection throughout
the star (Mullan \& MacDonald 2001; Feiden \& Chaboyer 2014) or
by blocking the emergence of radiative flux at the photosphere with
dark, magnetic starspots (e.g. Spruit \& Weiss 1986; MacDonald \& Mullan 2013; Jackson \&
Jeffries 2014a). Young stars on the pre-main-sequence (PMS) or at the zero-age main
sequence (ZAMS) are also highly magnetically active as a consequence of
their rapid rotation. If magnetic activity is responsible for inflating
the radii of fast-rotating binary components, then it seems likely that
the same phenomenon will be exhibited by low-mass PMS and ZAMS stars. 

\begin{figure*}
	\centering
	\begin{minipage}[t]{0.98\textwidth}
	\includegraphics[width = 180mm]{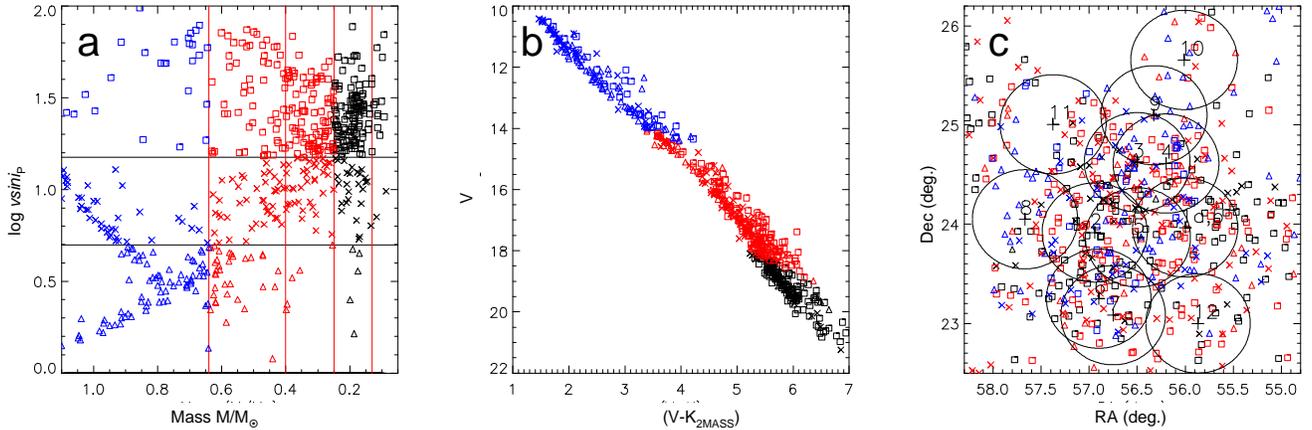}
	\end{minipage}
	\caption{Target selection: panel (a) shows $(v \sin
          i)_p$ as a function of estimated mass (see section 2.1) for low mass stars in the Peiades with measured period.
           Symbols correspond to the predicted observed $(v
          \sin i)_p$ (for an average spin-axis inclination and no
          radius inflation; triangles indicate $(v\sin i)_p < 5$
          km\,s$^{-1}$, crosses have $5< (v\sin i)_p <15$ km\,s$^{-1}$
          and squares $(v\sin i)_p \geq 15$ km\,s$^{-1}$), whilst the
          different colours correspond to intervals in mass inferred
          from the observed $K_{\rm 2MASS}$ magnitude and $V-K$ colour (blue for $M>0.64M_{\odot}$,
          red for $0.64>M/M_{\odot}>0.25$ and black for
          $M<0.25M_{\odot}$.  Panel (b) shows $V$ versus $V-K_{\rm 2MASS}$ colour magnitude diagram for the same data set using the same symbols
          and colour coding. Panel (c) shows 
          spatial distribution of low mass targets in the central
          region of the Pleiades. The large circles
          show the fields of view of the twelve observed WIYN Hydra
          configurations (see Table 1).}
	\label{fig1}
\end{figure*}

Substantial evidence has emerged that this is the case. Inflated
radii have been invoked to explain a number of puzzles: the
rotation-dependent anomalous
colours of PMS and ZAMS stars (Stauffer et al. 2003; Kamai et al. 2014;
Covey et
al. 2016); the rotation-dependent scatter of lithium depletion seen among PMS and ZAMS
stars of similar mass and age (e.g. Somers \& Pinsonneault 2014,
2015a,b) and the discrepancies between model predictions and the
measured masses, radii and luminosities of PMS and ZAMS eclipsing
binary systems (Kraus et al. 2015, 2016; David et al. 2016).  The
adoption of ``magnetic models'' for low-mass stars leads to the
inference of significantly older ages (by a factor of two) and higher
masses for PMS stars (Feiden 2016; Messina et al. 2016; Jeffries et
al. 2017) from the Hertzsprung-Russell diagram and 
commensurately longer empirically determined timescales for the
duration of star and planet formation.

Ideally, testing PMS/ZAMS magnetic models would involve direct measurements
of stellar masses and radii, but these are only directly accessible for
binary stars that might also be affected by tidal locking. Even the nearest
PMS/ZAMS stars are just too far away for precise interferometric radius
determinations. Indirect estimates of stellar radii can be made from
measured luminosities and $T_{\rm eff}$ determined from 
colours or spectroscopy. This approach was adopted by Somers \& Stassun
(2017) in the Pleiades cluster (age 120 Myr) and they found some
evidence for inflated radii (by 10--30 per cent) in ZAMS K-stars with
rotation periods $<2$ days compared with slower rotating stars of
similar spectral type. Unfortunately this technique is subject to
systematic errors in the $T_{\rm eff}$ estimation and is insensitive to
any inflation caused by dark starspots, since these can reduce the
luminosity of a star whilst leaving the colour and spectral type
largely unchanged.

An alternative geometric technique is to use the product of rotation
period ($P$ in days) with projected equatorial velocity ($v \sin i$ in
km\,s$^{-1}$), which yields the {\it projected} stellar radius, in solar units;
\begin{equation}
R\sin i = 0.0198\, P\, v \sin i\, 
\end{equation} 
(e.g. Rhode, Herbst \& Mathieu 2001; Jeffries 2007).
By assuming a random spin axis orientation (e.g. Jackson \& Jeffries
2010a) and taking account of
observational biases, a set of $R \sin i$ estimates can be modelled to
determine the true average radius for a group of stars, with a
precision that improves with larger samples.

The $P\,v\sin i$ method was used by Hartman et al. (2010) to
estimate average radii for G- and K-stars in the Pleiades using
rotation periods from the HATNet survey and $v \sin i$ measurements
from a variety of literature sources. They concluded that stars with
$M\geq 0.85 M_{\odot}$ have radii consistent with non-magnetic
evolutionary models, but that stars with $0.6<M/M_{\odot}<0.85$ were 10
per cent larger than predicted. Jackson \& Jeffries (2014a) used the
same dataset and a similar modelling technique to compare the radii of
Pleiades stars with $0.55 \leq M/M_{\odot}<1.0$ with the interferometric
radii of inactive main sequence field stars, finding an over-radius of
$13 \pm 3$ per cent. Lanzafame et al. (2017) performed their own
analysis showing that the effect appears to be driven by a $\sim 30$
per cent inflation for a subset of stars with $0.6 < M/M_{\odot}<0.8$
that are intermediate in rotation rate, but that faster rotators or
more massive stars have radii consistent with model predictions.

The goal of the present study is to extend these studies in the
Pleiades to lower masses. The motivation is twofold. First, Jackson et
al. (2009) and Jackson \& Jeffries (2014a) applied these techniques to
M-dwarfs in NGC 2516, a cluster with a similar age to the Pleiades,
finding a dramatic radius inflation that increased with decreasing
mass, reaching $\sim 40$ per cent for the lowest masses considered
($\simeq 0.25 M_{\odot}$). Whilst a number of systematic effects
(differential rotation, binarity) have been considered and accounted
for, it is {\it possible} that the rotation periods used, which were
based on a relatively short ground-based campaign, might have led to a
mistaken upward bias. The result has however been supported (with low
precision) by measurements of
a few M-dwarfs in an even younger cluster (NGC~2547, Jackson et
al. 2016), but we wish to confirm whether such large radius increases are
present in a larger sample with better-determined periods (see
below). Second, the predictions of the different flavours of magnetic
model differ for low-mass stars that are mostly or fully
convective. Magnetic inhibition of convection predicts a stronger
effect in higher mass stars with large radiative cores (Feiden \& Chaboyer
2015), whereas inflation due to starspots is predicted to be more
effective in fully convective stars, especially those which have yet to
reach the ZAMS, which is the case for stars with $M<0.4 M_{\odot}$ at the
age of the Pleiades (Jackson \& Jeffries 2014a). Hence measurements of
radius inflation across the ``fully convective boundary'' could be
diagnostic of the mechanism by which radius inflation occurs.

In this paper we present new results for low-mass stars in the
Pleiades. The cluster was included in the Kepler K2 mission (Howell et
al. 2014) for 72 days during ``campaign 4'', starting 8th February
2015. Rebull et al. (2016a) reported 760 rotation periods for low-mass
stars, including $\sim 600$ in the range $0.1<M/M_{\odot}<0.9$. To this
can be added a further 40 periods for low mass stars ($\leq 0.45
M_{\odot}$) from the Palomar Transit Factory survey (Covey et
al. 2016), and together these provide a large catalogue of reliable
rotation periods that bridge the fully convective boundary. 

We have
targeted these objects with fibre spectroscopy from the
WIYN\footnote{The WIYN Observatory is a joint facility of the
  University of Wisconsin Madison, Indiana University, the National
  Optical Astronomy Observatory and the University of Missouri.} 3.5-m telescope
at the Kitt Peak National Observatory, in order 
to determine $v \sin i$ and hence
distributions of $R \sin i$. In sections 2 and 3 we describe the target
selection and the measurements that were made at the WIYN
telescope. Sections 4 and 5 describe the analysis of these spectra to
determine $v \sin i$ for individual objects and to determine the
average over-radius for groups of objects. In section 6 we discuss the
significance of our results and compare them with the predictions of
non-magnetic models and models that include the effects of magnetic
inhibition of convection and starspots. Section 7 contains our
conclusions.
  
\section{Spectroscopic observations}

\subsection{Target selection}
Potential targets were selected from lists of Pleiades members with
measured periods reported by Rebull et al. (2016a), Covey et al. (2016)
and Hartman et al. (2010). Rotation period data were taken preferentially from
Rebull et al. (705 targets), then from Covey el al. (44) and lastly from
Hartman et al. (64). Data were matched with the 2MASS catalogue (Skrutskie el al. 2006)
 to give target co-ordinates (RA and Dec) and the apparent
$K_{\rm 2MASS}$ magnitude. Figure~1 shows the distribution of potential
targets in RA and Dec and in the colour-magnitude diagram. 
Targets for our fiber-spectroscopy study were selected from a
10 square degrees with the highest target density.

\begin{table*}
\caption{Hydra Configurations observed in the Pleiades. The positions
  are those of the field centres.}
\begin{tabular}{crcccccrccc} \hline
 Config. & File & Range of  & Date & UT of     &RA      & Dec   & Exposure & Number   & Fibres on & Fibres on \\
 number  & number& $I$ magnitude&    & exposure \#1 & \multicolumn{2}{c}{(J2000)} & time (s) & exposures & targets  & sky \\\hline
1a&    1013 &12.3 to 17.3 & 2016-09-24 & 08:07:09 & 03:43:59.99  & 23:57:59.94  & 3600 & 3 & 48 & 27\\
1b&    2046 &12.3 to 17.3 & 2016-09-25 & 07:59:09 & 03:43:59.99  & 23:57:59.94  & 3600 & 4 & 48 & 27\\
2&    4054 &12.3 to 17.3 & 2017-01-03 & 05:43:01 & 03:47:46.00  & 23:55:00.08  & 3600 & 6 & 52 & 20\\
3&    6063 &12.3 to 17.3 & 2017-01-05 & 04:45:10 & 03:45:59.99  & 24:37:29.93  & 3600 & 5 & 45 & 26\\
4&   12053 &~9.6 to 14.0 & 2017-01-17 & 05:34:48 & 03:44:47.99  & 24:36:44.96  &  600 & 6 & 22 & 20\\
5&   13025 &~9.6 to 14.0 & 2017-01-18 & 07:11:14 & 03:45:58.00  & 23:52:00.01  &  600 & 6 & 22 & 28\\
6&   14025 &~9.6 to 14.0 & 2017-01-19 & 02:03:00 & 03:47:35.99  & 23:14:59.98  & 3000 & 5 & 40 & 29\\
7&   21063 &~9.6 to 14.0 & 2017-02-02 & 02:20:17 & 03:46:59.99  & 23:04:59.96  & 1200 & 3 & 30 & 25\\
8&   21066 &~9.6 to 14.0 & 2017-02-02 & 03:47:40 & 03:50:39.99  & 24:03:00.06  & 1200 & 3 & 24 & 26\\
9&   21075 &~9.6 to 14.0 & 2017-02-02 & 05:53:13 & 03:45:18.10  & 25:05:58.01  & 1200 & 3 & 22 & 25\\
10&   22073 &~9.6 to 14.0 & 2017-02-03 & 02:25:01 & 03:44:02.79  & 25:39:22.08  & 1200 & 3 & 7  & 25\\
11&   22076 &~9.6 to 14.0 & 2017-02-03 & 03:51:08 & 03:49:30.00  & 25:00:30.05  & 1200 & 3 & 18 & 26\\
12&   22079 &12.3 to 17.3 & 2017-02-03 & 05:19:13 & 03:43:27.99  & 23:00:00.05  & 3600 & 2 & 33 & 25\\\hline           
    \end{tabular}
  \label{observations}
\end{table*}

\begin{table*}
\caption{Properties of observed science targets in the Pleiades and reference
  slow rotators in Praesepe. Masses and radii are estimated from the
  models of BHAC15. The final column gives the
  predicted equatorial velocity -- see section 2.1. A sample of the
  table is given here, the full table is made available electronically.}
\begin{tabular}{lccccccccccc} \hline
Target name		&RA	&Dec	&$K_{\rm 2MASS}$&$V-K$
&Period  &Ref. &$BC_K$	&$\log L/L_{\odot}$ & $M/M_{\odot}$
&$R/R_{\odot}$ &$(v \sin i)_p$\\
(2MASS)		&\multicolumn{2}{c}{(J2000)}		  & (mag)   	&(mag)&P (days)&   *   &(mag)&&& &(km\,s$^{-1}$)\\\hline
J03414895+2303235	&	03 41 48.951	&	+23 03 23.54	&	13.19	&	6.09	&	0.239	&	1	&	2.86	&	-2.26	&	0.19	&	0.24	&	39.3	\\
J03415671+2358434	&	03 41 56.716	&	+23 58 43.42	&	13.25	&	5.76	&	0.401	&	1	&	2.82	&	-2.27	&	0.18	&	0.24	&	23.3	\\
J03415864+2257020	&	03 41 58.648	&	+22 57 02.00	&	11.90	&	4.78	&	6.842	&	1	&	2.72	&	-1.68	&	0.40	&	0.38	&	2.2	\\
J03421789+2406578	&	03 42 17.890	&	+24 06 57.83	&	12.97	&	5.53	&	0.603	&	1	&	2.80	&	-2.15	&	0.22	&	0.26	&	17.0	\\
J03422626+2351386	&	03 42 26.266	&	+23 51 38.67	&	13.45	&	5.97	&	0.496	&	1	&	2.85	&	-2.36	&	0.17	&	0.22	&	17.7	\\
J03422941+2247261	&	03 42 29.418	&	+22 47 26.19	&	10.92	&	4.11	&	0.325	&	1	&	2.62	&	-1.25	&	0.56	&	0.52	&	62.4	\\
J03423396+2411008	&	03 42 33.960	&	+24 11 00.81	&	13.42	&	5.89	&	0.564	&	1	&	2.84	&	-2.34	&	0.17	&	0.23	&	15.8	\\
J03424184+2400158	&	03 42 41.848	&	+24 00 15.81	&	12.68	&	5.17	&	0.671	&	1	&	2.76	&	-2.01	&	0.26	&	0.29	&	17.1	\\
J03424239+2320218	&	03 42 42.396	&	+23 20 21.87	&	11.45	&	5.28	&	0.269	&	1	&	2.77	&	-1.53	&	0.46	&	0.43	&	63.4	\\\hline
\multicolumn{10}{l}{* Period measurement taken from (1) Rebull et al. 2016a, (2) Covey et al. 2016, (3) Hartman et al. 2010, (4) Douglas et al. 2017}\\    \end{tabular}
  \label{targets}
\end{table*}

The targets were prioritised according to their mass, $M$ (highest
priority was given to lowest masses, but with a practical faint magnitude limit
of $I=17.3$), and a
prediction of their observed projected equatorial velocity, $(v \sin
i)_p =(\pi /4) 50\,R/P$ in km\,s$^{-1}$, where $P$ is the rotation
period in days and $R$ is the estimated stellar radius in solar
units. Masses and radii were estimated by comparing 
the luminosity of the potential target with the predictions of a
120\,Myr solar metallicity, {\it non-magnetic} model isochrone from
Baraffe et al. (2015) (hereafter BHAC15). The factor of $\pi/4$ is the mean
value of $\sin i$, based on a crude assumption of an unbiased, random distribution of rotation
axis orientations.  The adopted metallicity and age are consistent with
[Fe/H]$=0.03 \pm 0.05$ reported by Soderblom et al. (2009) and the
lithium depletion boundary age of $125\pm 8$\,Myr given by Stauffer,
Schultz \& Kirkpatrick (1998). The luminosity of each source was estimated from its $M_K$
value (accounting for extinction and distance, see below) using a bolometric
correction calculated using $(V-K)_0$ and the BHAC15 models.  An
intrinsic distance modulus of $5.67\pm 0.02$\,mag was assumed (Melis et
al. 2014), a conversion of $K_{\rm CIT}=K_{\rm 2MASS}+0.024$ (Carpenter
2001) and a reddening of $E(B-V)=0.032$\,mag (An, Terndrup \&
Pinsonneault 2007). Using the relations of Rieke \& Lebofsky (1985)
this reddening corresponds to extinction of $A_V=0.10$\,mag and
$A_K=0.01$\,mag.

The targets were binned according to $(v \sin i)_p$ (see
Fig.~1a). Highest priority was given to the faster rotating targets
with $(v \sin i)_p \ge 15$km\,s$^{-1}$, which are expected to yield
measurable $v\sin i$ values at the resolution of the spectroscopic data
(see section 2.2). Second priority was given to targets with
$(v \sin i)_p \le 5$\,km\,s$^{-1}$ to provide a sample of very slow
rotators which are expected to have unbroadened spectral lines and
serve as a baseline for calibrating the measured $v \sin i$ as a function
of spectral line width broadening (see section 3.4).

\subsection{Observations}

The WIYN Hydra multi-object spectrograph (Bershady et al. 2008)
consists of a robotic positioner that can position up to 83 fibres,
each with a 3 arcsecond diameter (we used the ``blue'' fibre cable),
across a 1 degree diameter unvignetted field of view at the Nasmyth
focus of the 3.5-m WIYN telescope. The fibers were used in conjunction
with the bench spectrograph, an echelle grating and an order-sorting
filter to provide spectra with a resolving power of $\simeq 17,000$. An
STA1 2600$\times$4000 pixel CCD camera was used in $2\times2$ binning
mode to record spectra covering $\sim 400$\AA\ centred at $\sim
7850$\AA. The FWHM of a resolution element corresponded to about 2.5
binned pixels. Twelve field centres were chosen to maximise the number of high
priority targets (see section 2.1). Spare fibres were allocated to
second priority targets and to repeat observations of targets in cases
of over-lap between fields. Finally, $\sim$20 spare fibres were
allocated to clear sky $>$20 arcsec away from the nearest source in the
2MASS catalogue.

The observing program was performed over 9 nights during a 6 month
period from September 2016 to February 2017, although poor weather
restricted the total observing time available. Details of when each of
the twelve fields was observed (field 1 was observed on two nights),
how many targets were in each field and how long the fields were
exposed for are given in Table 1. The range in apparent $I$
magnitude (the relevant magnitude for the wavelength at which we
observed) within a particular configuration was restricted to $<5$
magnitudes to limit any cross-contamination of spectra between adjacent
fibres. To make best use of
varying observing conditions we further divided the targets into
``bright'' and ``faint'' samples (with overlap). For configurations of
fainter targets ($12.3<I<17.3$), several one hour exposures were
required to produce sufficient signal to noise (SNR) in the spectra to
allow resolution of $v\sin i$ in the faintest targets. The brighter
targets ($9<I<14$) required less time and could be observed more
readily in partially cloudy skies. The names, co-ordinates, photometry,
rotation periods, estimated masses and radii (from the BHAC15
models) and derived luminosities for the 324 individual Pleiades targets that were actually
observed are listed in Table 2.

\subsection{Additional observations of slowly rotating stars}

There were an insufficient number of slow-rotating M-dwarfs in the Pleiades to
adequately characterise the width of spectral lines in stars with
negligible rotational broadening (see section 3.4). To that end, also
listed at the end of Table 2, are 88 low-mass targets from Praesepe, an
older cluster, which also has measured periods based on K2 observations
(Douglas et al. 2017) and which contains a higher proportion of
slow-rotating M dwarfs 
($(v \sin i)_p \le 3\,$km\,s$^{-1}$).  The spectra of these stars were
observed on the same nights as the Pleiades targets and with exactly
the same Hydra spectrograph set up and were taken from a comprehensive
set of observations made of low-mass Praesepe stars with known periods,
which will be reported on in a subsequent paper.  $M_K$ values are
estimated assuming a distance modulus of 6.29$\pm$0.07 mag (van Leeuwen
2009) and zero reddening. Stellar masses and radii were estimated from
intrinsic $V-K$ colour and the BHAC15 models, assuming a cluster age of
670\,Myr (e.g. Cummings et al. 2017).

\section{Data reduction}
Many of the target spectra were faint, requiring an optimal extraction
strategy to provide sufficient signal-to-noise ratio (SNR) for useful
analysis. Strong sky emission lines were a dominant feature in the
fainter spectra. For these reasons we used purpose-built software for
data reduction based on the pipeline described in Jackson \& Jeffries
(2010b), adapted where necessary, to the characteristics of the WIYN
telescope and Hydra spectrograph.

\subsection{Extraction of target spectra}
Images of the science fields and associated flat and arc exposures were
debiased and rebinned to compensate for the initial curvature of the
spectra on the CCD image. The flat frames were the median of 11
tungsten-lamp flat exposures recorded in the afternoon prior to
night-time observations. One dimensional spectra were optimally
extracted from the science frames using the procedure described by
Horne (1986). Counts per bin and uncertainties were calculated for a
gain of 0.44 electrons/ADU and a read out noise of 3.1 electrons.

\begin{figure}
	\centering
	\includegraphics[width = 85mm]{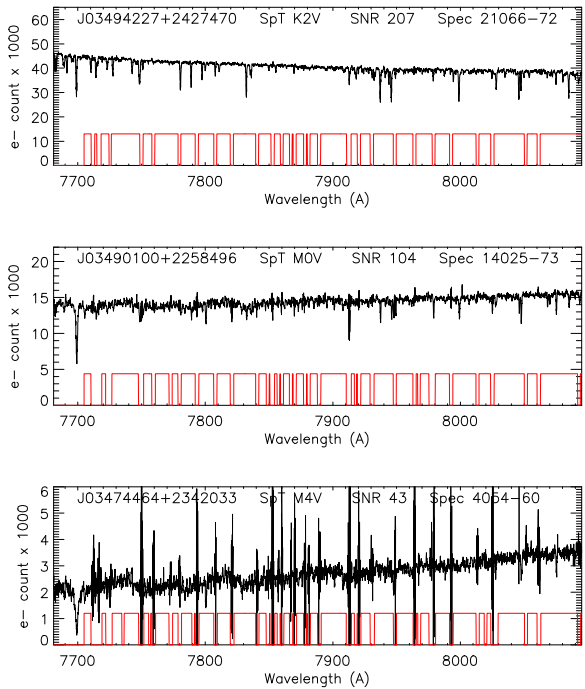}
	\caption{Representative spectra of stars in the Pleiades
          showing the presence of skylines in lower SNR spectra. 
The spectral types (SpT) shown are estimated from ($V$-$K$)$_0$ using 
the relationship proposed by Kenyon \& Hartmann (1995). The lower (red) 
plots indicate sections of the spectra that are masked to minimise the 
effect of sky lines for measurement of $RV$ and $v\sin i$. }
	\label{fig2}
\end{figure}

Arc spectra were extracted from Thorium-Argon lamp exposures recorded
during the day prior to observations. Gaussian fits were used to
determine the locations of 6 well-spaced unsaturated arc lines 
recorded through each fibre. Cubic polynomial fits to these
were used to rebin spectra onto a common wavelength scale. The observations
in September 2016 were centered on $\sim$7830\AA\ with a common
wavelength range of 7620--8035\AA. Subsequent observations were centered on
$\sim$7890\AA~\ with a common wavelength range of 7681--8095\AA. A
fine adjustment was made to the wavelength scale applied to each
observation to compensate for drift between day-time
calibration and night-time observations. The adjustment was determined
by comparing the measured wavelength of six strong, unblended emission
lines in the median sky spectra with their reported wavelengths. The
weighted mean skyline correction varied from $-0.6$ to $+1.5$~km\,s$^{-1}$
($-0.02$\AA~to $+0.04$\AA) with, in most cases, an uncertainty of $\le$
0.25~km\,s$^{-1}$ although a higher uncertainty (0.9~km\,s$^{-1}$) was
found for configuration 5 (see Table 1).

Target spectra were sky subtracted using fibre efficiencies estimated
from the amplitude of the flat field spectra which, when checked,
showed good agreement ($\sim$1.5 per cent rms) with the throughput
measured from a twilight sky exposure of the same configuration on the
same night. Spectra from repeated exposures in the same configuration (see
Table 1) were corrected for heliocentric radial velocity and the median
taken to produce final spectra, and corresponding uncertainties, in
0.1\AA~steps over the common wavelength ranges. Figure~2 shows
typical spectra with spectral types estimated from ($V$-$K$)$_0$
(Kenyon \& Hartmann, 1995). Despite care
taken with sky subtraction, lower SNR spectra show residual sky lines
which could adversely effect $v\sin i$ measurements if not masked
prior to further analysis. A total of 411 spectra were collected for
324 separate targets in the Pleiades. 10 spectra were rejected from further
analysis because of a low SNR ($\leq 9$) reducing the number of Pleiades targets to 319.

\subsection{Measurement of radial velocities and spectral broadening}
$RV$ and $v\sin i$ were measured by cross
correlating the median spectra of individual targets with the spectra of
standard stars and then fitting a Gaussian function
to characterise the peak in the cross-correlation function (CCF). 
$RV$s were determined from the position
of the peak in the CCF and the spectral broadening was estimated from
the increase in full width half maximum (FWHM) of the Gaussian fit
with respect to FWHM$_0$, the CCF FWHM measured for slow-rotating
stars of similar spectral type. 

For this analysis the spectra were truncated
shortward of 7705\AA~to avoid strong telluric features. Spectra were also masked at the positions of the
strong skylines (see Fig.~2) and rebinned (with 10000 points) on a logarithmic
wavelength scale, The masked spectra were cross-correlated with template
spectra taken from the UVES atlas (Bagnulo et al. 2003). Five templates 
were used to approximately match the expected spectral types of
Pleiades targets based on their (binned) absolute K magnitude (see Table 3). Cross-correlation yielded $RV$ and FWHM values
for 319 targets, but 5 of these were rejected after visual inspection;
2 were well-separated spectroscopic binaries and 3 had very poor
Gaussian fits to the peaks in their CCFs.
\begin{table}
\caption{Calibration standards. Target spectra are cross correlated 
with spectra of CCF templates (see section 3.2). Spectra of $v\sin i$ standard stars
are used to define calibration curves of $v\sin i$ versus FWHM (see section 3.3).}

    \begin{tabular}{llll} \hline
No. &$M_K$  & CCF template              &$ v\sin i$\\
 & range    & (spectral type)     &  standard(s)\\\hline
1 & $>$5.5	  & HD~34055	(M6V)	      &	Gl\,133/Gl\,285\\
2 & 4.9 - 5.5 & HD~130328 (M3III)	  &	Gl\,133/Gl\,285\\
3 & 4.4 - 4.9 & HD~156274	(M0V)	    &	Gl\,184/Gl\,205\\
4 &3.9 - 4.4	& HD~10361	  (K5V)	    &	Gl\,184/Gl\,205\\\hline
    \end{tabular}
  \label{standards}
\end{table}

\subsection{Measurement precision}
The precision of $RV$ and FWHM measurements were determined empirically
from the change in $RV$ and FWHM between repeated measurements of the same
target either in the same configuration (1a
and 1b in Table 1) or for targets present in two or more configurations. To
maximise the sample size, targets from Praesepe (see section
2.3) were also included in the analysis to give a total of 174 repeats compared
with 65 in the Pleiades alone. Assuming that the standard deviation of
the measurements of both $RV$ and $v \sin i$ are proportional to the
FWHM of the CCF, the distribution of measurement uncertainties measurement
was characterised by a t-distribution, with $\nu$ degrees of freedom,
with a width that is scaled by a function that features a fixed systematic component plus a component that
depends on SNR. In the limit of $\nu \rightarrow \infty$ this would be
equivalent to a Gaussian with a standard deviation given by the scaling function.

Uncertainties were estimated from repeat observations of individual targets
($E_{RV}$=$\Delta RV/\sqrt{2}$~and
$E_{\rm{FWHM}}$=$\Delta$FWHM/$\sqrt{2})$. The distributions of these
were modelled in order to choose an
appropriate $\nu$ and to obtain empirical values
for the dimensionless parameters, $A$,$B$,$\alpha$ and $\beta$ of the scaling functions
$S_{RV}$ and $S_{\rm{FWHM}}$, where $S$ in each case is a measure of
the standard deviation and defined as
\begin{equation}
    S_{RV} = \rm{FWHM}\sqrt{A^2 + (B/SNR)^2}\, ,  
\end{equation}
where $A=0.025$ and $B=0.95$, and
\begin{equation}
    S_{\rm{FWHM}} = \rm{FWHM}\sqrt{\alpha^2  + (\beta/SNR)^2}\, , 
\end{equation}
where $\alpha=0.036$ and $\beta =0.68$. Given that the FWHM is $\geq 22$ km\,s$^{-1}$, this indicates
absolute uncertainties in $RV$ of at least 0.5 km\,s$^{-1}$ and FWHM
uncertainties of at least 0.8 km\,s$^{-1}$, once the SNR greatly
exceeds $\sim 40$.

The upper plot in Fig.~3 shows the cumulative distribution of
the normalised uncertainty in $RV$ (i.e the ratio of measured
uncertainty to the uncertainty predicted by $S_{RV}$  using the best fit values of
$A$ and $B$. A t-distribution
with $\nu=4$ degrees of freedom is a good match to the data. Note that
a finite value of $\nu$ indicates that the tails of the distribution
are more prominent than a normal distribution and that a 68.3 per cent
confidence interval would be given by $1.14\, S_{RV}$ for $\nu=4$.
The lower plot shows the distribution of normalised 
measurement precision for FWHM. In this case the tail of
the normalised distribution is slightly more extended such that a 
t-distribution with $\nu=3$ provides a better fit and a 68.3 per cent
confidence error bar would be given by $1.20\, S_{\rm FWHM}$.

\begin{figure}
	\centering
	\includegraphics[width = 70mm]{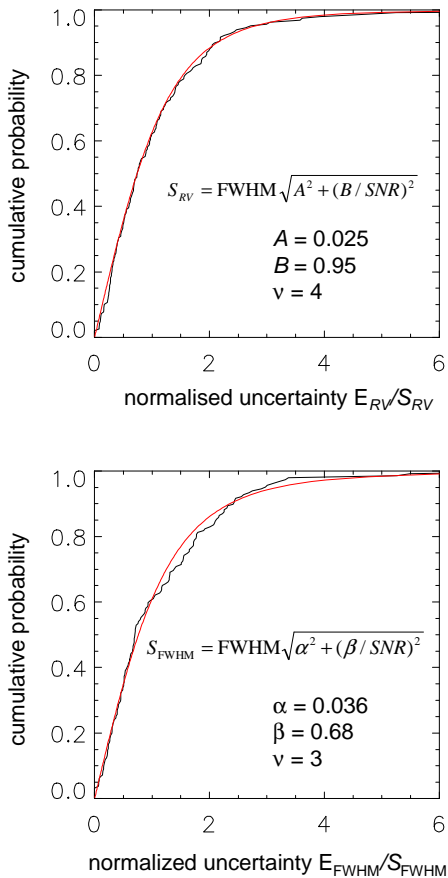}
	\caption{Measurement precision of $RV$ and FWHM estimated from
          the the CCF with template field stars (see Table~3): Plots
          shows the cumulative distribution functions of the
          uncertainties (normalised using equations~2 and~3 with
          the parameters shown on the plots).}
	\label{fig3}
\end{figure}

\begin{table*}
\caption{Measured values of relative $RV$, FWHM, $v\sin i$ and $R\sin i$. When the relative uncertainty in $v\sin i$ 
is greater than 30 per cent an upper limit of $v\sin i$ is shown based on the measurement uncertainty in FWHM (see section 3.4.3). 
The corresponding upper limit in $R\sin i$ is treated as left-censored data in the maximum likelihood analysis determination of over-radius. A sample of the Table is shown here, the full version is available electronically.}
\begin{tabular}{lrrrrrrrrrrrr} \hline
Target name &	$M_K$ &$\log L/L_{\odot}$&Period
&SNR&$RV$&$S_{RV}$&FWHM&$S_{\rm FWHM}$& FWHM$_0$&$v\sin i$&$S_{v\sin i}$&$R\sin i$\\
(as 2MASS)  &(mag)&      &(d)	&   &(km/s)&(km/s)&(km/s)&(km/s)&(km/s)&(km/s)&(km/s)&($R_{\odot}$) \\\hline		
J03414895+2303235	&	7.51	&	-2.26	&	0.239	&	9.9	&	-1.2	&	2.9	&	29.2	&	2.3	&	24.2	&	18.2	&	4.1	&	0.087	\\
J03415671+2358434	&	7.57	&	-2.27	&	0.401	&	15.9	&	0.7	&	2.4	&	37.6	&	2.1	&	24.2	&	28.4	&	2.2	&	0.228	\\
J03415864+2257020	&	6.22	&	-1.68	&	6.842	&	36.3	&	-1.2	&	0.9	&	24.7	&	1.0	&	24.2	&	$<$10.6	&	 ---	&	$<$1.44	\\
J03421789+2406578	&	7.29	&	-2.15	&	0.603	&	47.7	&	0.1	&	0.9	&	26.9	&	1.0	&	24.2	&	13.2	&	2.5	&	0.159	\\
J03422626+2351386	&	7.77	&	-2.36	&	0.496	&	13.7	&	-1.1	&	1.9	&	25.4	&	1.6	&	24.2	&	$<$12.8	&	 ---	&	$<$0.13	\\
J03422941+2247261	&	5.24	&	-1.25	&	0.325	&	73.7	&	3.1	&	2.1	&	74.8	&	2.8	&	24.7	&	50.8	&	1.4	&	0.330	\\
J03423396+2411008	&	7.74	&	-2.34	&	0.564	&	28.5	&	-0.5	&	1.2	&	27.9	&	1.2	&	24.2	&	15.9	&	2.6	&	0.179	\\
J03424184+2400158	&	7.00	&	-2.01	&	0.671	&	65.3	&	-0.5	&	0.9	&	30.4	&	1.1	&	24.2	&	20.2	&	1.9	&	0.271	\\
J03424239+2320218	&	5.77	&	-1.53	&	0.269	&	45.7	&	0.6	&	1.1	&	33.4	&	1.3	&	24.2	&	23.8	&	1.7	&	0.128	\\\hline
\end{tabular}
  \label{vsini}
\end{table*}

\subsection{$RV$ and $v\sin i$ for the Pleiades targets}
Table 4 gives the measured $RV$ and FWHM and estimated
uncertainties of the 314 Pleiades targets with well defined CCFs that
can be used to determine stellar $v\sin i$. Where repeated measurements
were made of the same target the values shown in Table~4 are the weighted (by $S^{-2}$)
mean values.

\subsubsection{Cluster $RV$s}
The $RV$s in Table 4 are measured relative to the central $RV$ of the
cluster such that $RV_{\rm rel}=RV-RV_0$ where $RV_0$ is the median value
of the target $RV$s~measured relative 
to a particular CCF standard (see Table~3).
The dispersion of the measured $RV_{\rm rel}$, estimated from the
median absolute dispersion (MAD) of the target $RV$s,  is
$\sigma_{t}$=1.2\,km\,s$^{-1}$ (using the approximate relation
$\sigma_{t}=$MAD/0.68 for a t-distribution with $\nu=4$). $\sigma_{t}$
is due to the
combined effect of (a) intrinsic dispersion in the cluster, (b)
measurement uncertainties and (c) the effects of binarity. It is used
here to define a window of acceptable $RV$s for Pleiades
membership as $|RV_{rel}|<10$\,km\,s$^{-1}$. Using this criterion
eliminates 9 targets as possible non-members or short period
binary systems.

\begin{figure}
	\centering
	\includegraphics[width = 85mm]{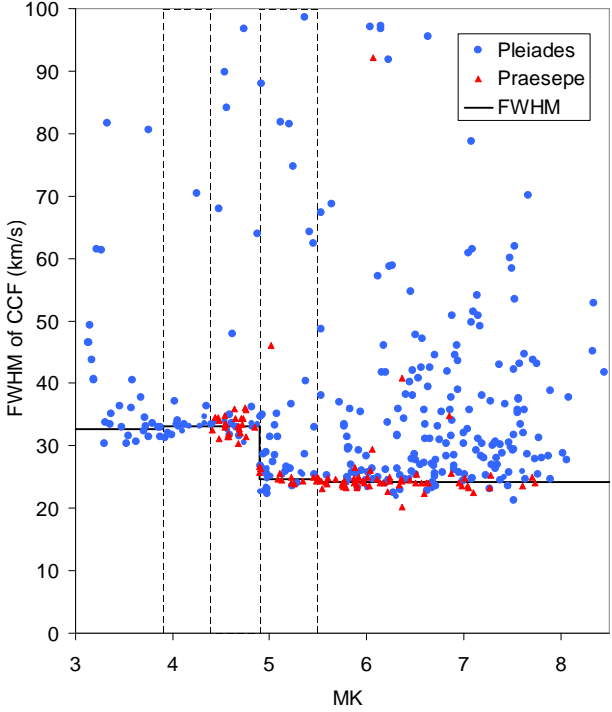}
	\caption{FWHM of the CCF measured on targets in the Pleiades
          (blue circles) and Praesepe (red triangles) as a function of
          $M_K$. Dashed vertical lines delineate the ranges where
          particular templates were used to calculate the CCF (see
          Table 3). Horizontal bars show the derived zero-point,
          FWHM$_0$, for slow-rotating stars as a function of $M_K$.}
	\label{fig4}
\end{figure}

\begin{figure}
	\centering
	\includegraphics[width = 77mm]{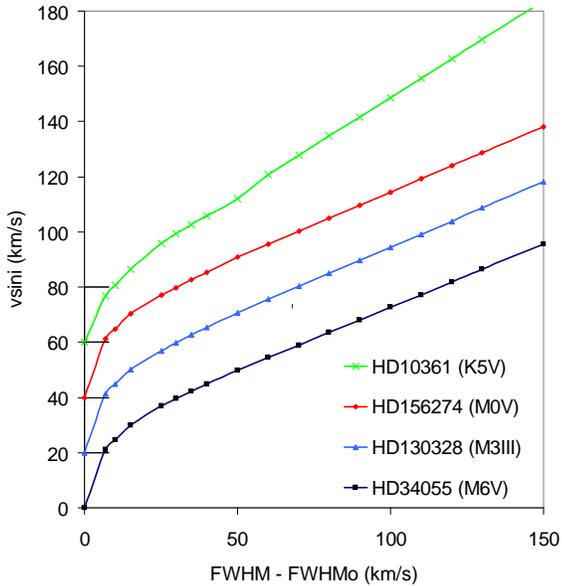}
	\caption{Calibration curves of $v\sin i$ as a function of the FWHM of the CCF. Results are shown for the four CCF templates listed in Table~3 cross correlated with artificially broaded vsini standards of similar spectral type (see section 3.4.3). Curves are offset on the vertical axis in increments of 20\,km\,s$^{-1}$.}
	\label{fig17}
\end{figure}
  
\subsubsection{The FWHM zeropoint}
Figure~4 shows FWHM of the CCFs as a function of $M_K$ for the Pleiades targets,
with vertical dashed lines showing the absolute magnitude ranges used to decide which
templates were used to calculate the CCF. Also shown are the FWHM
values for slow-rotating targets in Praesepe (i.e. targets with
$(v\sin i)_p<3$\,km\,s$^{-1}$). The FWHM of 93 (predicted) slowly rotating
Praesepe targets together with 22 similar slow rotators in the Pleiades
were used to define the median CCF width for slow-rotating stars, FWHM$_0$, in
the respective $M_K$ bins. The relationship between the
FWHM$_0$ and $M_K$ is shown in Fig.~4 and given for each star in Table~4.

\subsubsection{$v \sin i$ values and their precision}
The target $v\sin i$ values were determined from
$\Delta$FWHM=FWHM-FWHM$_0$ using calibration curves produced by
artificially broadening the spectra of bright, slowly rotating standard
stars measured at the beginning and end of each observing night. The
$v\sin i$ standards used to calibrate each range of M$_K$ are given in
the final column of Table 3. The broadening kernel assumed a linear
limb darkening coefficient of 0.6 (Claret, Diaz-Cordoves \& Gimenez
1995). Figure~5 shows the relationship between $v \sin i$ and
$\Delta$FWHM for the range of spectral types in our sample. For values
of $v \sin i > 60$ km\,s$^{-1}$ we found that the exact calibration
depended on which template star was chosen, even at the same spectral
type.  Instead, for these rapid rotators, 
the relationship was linearly extrapolated from smaller
values, which we found provided a good match to the cross-correlation
FWHM values obtained from high SNR spectra of slowly rotating Pleiades
members of similar absolute magnitude that were artificially broadened.
This comparison also reveals that the calibration uncertainties appear
to grow from very small values at low $v \sin i$ to $\sim \pm 5$ per
cent for $v \sin i \geq 70$ km\,s$^{-1}$. However, since fewer than 5
per cent of the sample used to determine the over-radius in Pleiades
stars (see section 4) are in this regime, this systematic calibration
uncertainty leads to $<1$ per cent uncertainty in our final results and
we neglect it in the rest of the analyses.

The calibration curves in Fig.~5 vary approximately as $(v\sin i)^2 \propto
\Delta$FWHM for $v \sin i <60$ km\,s$^{-1}$  
and the precision in $v\sin i$ varies as $S_{v\sin i}=\tfrac{\partial v\sin i}{\partial \rm{FWHM}}S_{\rm{FWHM}}$. Using these expressions the \textit{relative} precision in $v\sin i$, defined here as  $E_{v\sin i} = \Delta v\sin i/(\sqrt{2}v\sin i)$, scales as
\begin{equation}
	\frac{S_{v\sin i}}{v\sin i} \simeq \frac{\sqrt{\alpha^2  +
            (\beta/SNR)^2}}{2(1 - \rm{FWHM}_0/\rm{FWHM})}\, .
\end{equation}
This formulation should be reasonably accurate up to $v \sin i \sim 60$
km\,s$^{-1}$, but may underestimate the uncertainties for the small
number of very fast rotators in our sample.
Figure~6 shows the variation of the relative uncertainty in $v\sin i$ with
FWHM/FWHM$_0$ for increasing levels of SNR. These plots were used to
calculate the $v \sin i$ uncertainty and also to define 
a threshold for FWHM/FWHM$_0$, as a function of SNR, that marks where
the level of rotational broadening is large enough to yield a resolved
value of $v\sin i$. We choose this threshold such that the relative
uncertainty in $v \sin i$ is $<0.3$ and targets with FWHM/FWHM$_0$  
below this level are assigned an upper limit value of $v\sin i$ at the value
corresponding to this threshold value. 
The threshold varies from  FWHM/FWHM$_0 = 1.12$ for
SNR$=10$ to FWHM/FWHM$_0 = 1.06$ for SNR$=100$. These threshold FWHM values
correspond to $v \sin i$ upper limits of $<13$ km\,s$^{-1}$ and $<10$ km\,s$^{-1}$
respectively, with a small dependence on spectral type.
  
\begin{figure}
	\centering
	\includegraphics[width = 80mm]{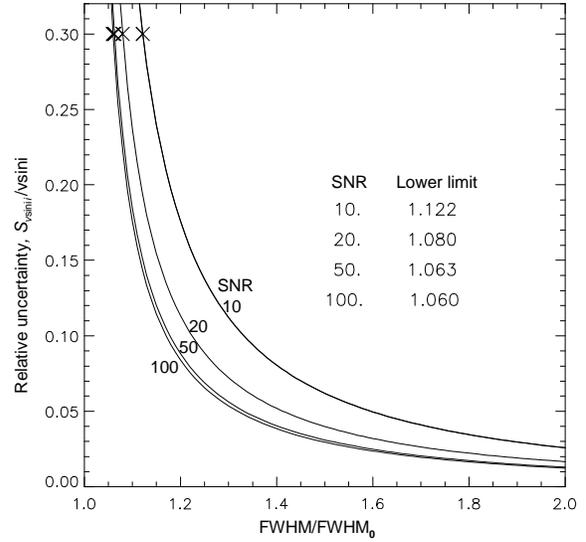}
	\caption{Measurement precision of $v\sin i$. The plot shows
          shows the relative uncertainty in $S_{v\sin i}/v\sin i$  (see
          eqn.~4) as a function of the measured FWHM for increasing
          levels of SNR. Tabulated values on the plot show the minimum
          levels of FWHM/FWHM$_0$ at each SNR value that can
          be discerned to yield a 30 per cent uncertainty in $v\sin i$.}
	\label{fig5}
\end{figure}

\subsubsection{Comparison of $v\sin i$ with other work}
The empirical analysis described above gives only a partial estimate of
the absolute accuracy, which is due to both the measurement
precision and the uncertainty in the absolute calibration. To assess
the calibration accuracy, our values for $v\sin i$ were compared with
those reported in the literature. Matches were found for
31 targets with $|RV_{rel}|\le10$\,km\,s$^{-1}$ and relative uncertainty in $v \sin i$ $\le$~0.3. 
The comparison is shown in Fig.~7 and demonstrates satisfactory agreement between the two datasets. 
There are three distinct outliers, two of which, marked (a) and (c) in Fig.~7, are identified as
spectroscopic binaries in Mermilliod et al. 1992. Linear regression of the two datasets (excluding the three outliers) shows no significant systematic difference between our $v\sin i$ measurements and the literature values.

\begin{figure}
	\centering
	\includegraphics[width = 75mm]{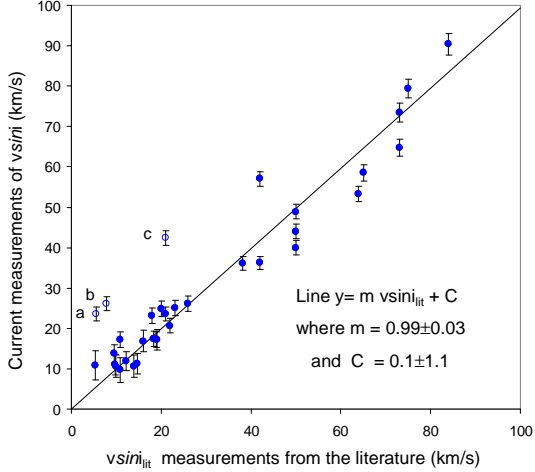}
	\caption{A comparison of $v\sin i$ values (with fractional
          uncertainties of $<$\,30 per cent 
	       and $|RV_{\rm rel}|$$\le$10\,km\,s$^{-1}$) with literature values (Queloz et al. 1998, 
	       Marilli et al. 1997, Stauffer and Hartmann 1987, Soderblom et al. 1993, O'Dell et al. 1994 and  Krishnamurthi et al.                1998). The solid line is a linear regression, neglecting the three outliers J03484894+2416027, J03434841+2511241 and                J03475973-2443528 marked as a,b and c respectively on the plot. Error bars are 68 per cent empirical                                 uncertainties in our measurement precision.}
	\label{fig6}
\end{figure}

\begin{figure}
	\centering
	\includegraphics[width = 85mm]{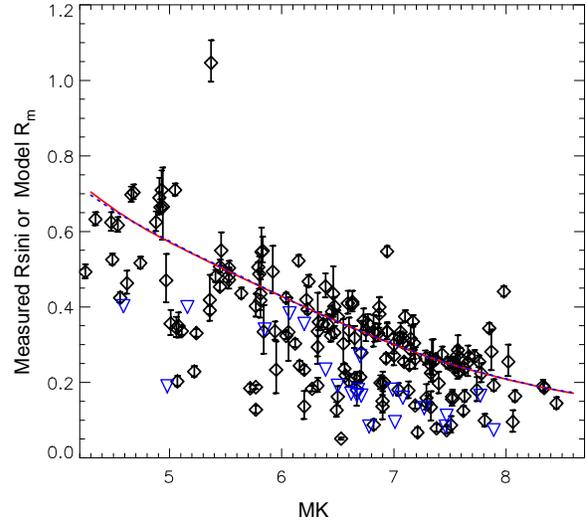}
	\caption{Variation of the projected radii, $R\sin i$ with M$_K$ for Pleiades K- and M-dwarfs. Diamonds with error bars show targets with a relative uncertainty in $R\sin i$$\le$~30 per cent. Triangles show upper limit values for targets with higher levels of uncertainty. Solid and dashed lines show predicted radii in solar units of the BHAC15 and Dartmouth (Dotter et al. 2008) 120\,Myr solar metallicity isochrones.}
	\label{fig7}
\end{figure}

\begin{figure*}
	\centering
	\includegraphics[width = 180mm]{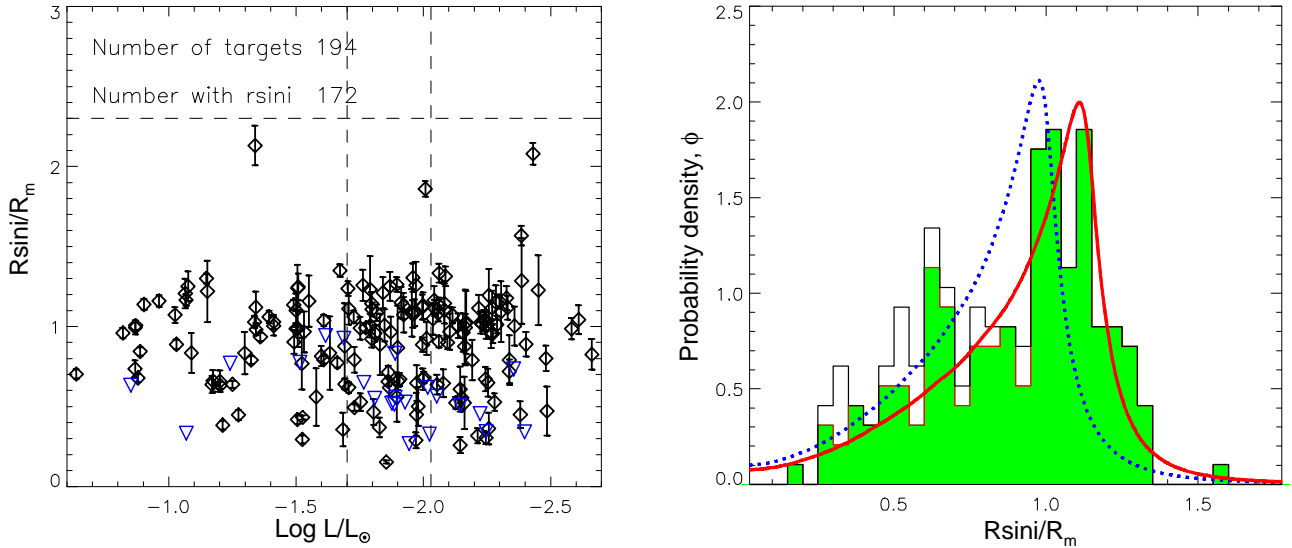}
	\caption{Normalised radii in the Pleiades. Plot A
          shows $r\sin i$ as a function of luminosity (see Table
          2). Diamonds with error bars show $r\sin i$ for targets with a
          relative uncertainty $\le$30 per cent. Triangles show upper
          limits for targets $(v\sin i)_p  >10$~km\,s$^{-1}$ with higher levels of
          uncertainty ($\ge 30$ per cent). Plot B shows a the number
          density ($P_{r\sin i}$) of targets with a relative
          uncertainty  $\le 30$ per cent as a solid histogram, with the
          open histogram including the stars with upper limits at their
          upper limit values. The dotted line shows
          $P_{r\sin i}$  for stars with the radii that are predicted by
          the BHAC15 evolutionary model (i.e. an over-radius
          $\rho=1$). The solid red line shows the maximum likelihood
          model which corresponds to an average over-radius of
          $\rho \sim 1.14$ relative to the BHAC15 model.}
	\label{fig8}
\end{figure*}

\section{Comparison of measured radii with current evolutionary models}
The measurements of $v\sin i$ in Table 4 are used with the reported
rotation periods to estimate projected radii $R\sin i$ (using Eqn. 1) for
Pleiades members with $|RV_{\rm rel}|<10$\,km\,s$^{-1}$. The uncertainty in $R\sin i$ is
estimated on the basis that the uncertainty in $v\sin i$ is much
greater than the uncertainty in period giving a {\it fractional}
uncertainty in $R\sin i$ of $S_{v\sin i}/v\sin i$, as in Eqn.~4. 
For targets where this fractional uncertainty is greater than 0.3, we assign
upper limits to $R\sin i$ as calculated from the target's period
and the upper limit to $v\sin i$. 

In the subsequent analyses the sample is restricted to those objects
with $(v \sin i)_p > 10$ km\,s$^{-1}$. This threshold is chosen to
approximately match the resolution limit of our measurements of $v \sin
i$. Whilst our analyses {\it do} incorporate upper limits, enlarging
the sample to include more slowly rotating stars simply adds many upper
limits that do not constrain any over-radius and just add more noise to
the results. The reader should then be aware that our radius
measurements can only apply to those relatively fast-rotating stars
that are in this restricted sample. The effects of increasing the $(v
\sin i)_p$ threshold to a more restrictive 15 km\,s$^{-1}$ (and
decreasing the sample size) has no systematic influence on the results (see section 5.1).
 
Figure 8 shows $R\sin i$ versus $M_K$ for the restricted
sample together with predicted model radii $R_m$
from the BHAC15 and Dartmouth evolutionary codes (Dotter
et al. 2008). For practical purposes these two models are identical over the
range of the data considered here. In what follows, the ratio of projected radius to
model radius at the target luminosity,
$r\sin i = R\sin i/R_m$, is referred to as the ``normalised radius''. In the absence of radius inflation the
distribution of $r\sin i$ would simply reflect the values of $\sin i$ in the
sample convolved with any measurement uncertainties and biases. Where
targets show radius inflation (i.e. $R/R_m>1$) then the $r\sin i$
distribution would also be scaled by a similar amount.

Fig.~9 recasts Fig.~8 to show $r\sin i$ as a function of $\log
L/L_{\odot}$ for targets with $-2.7 \leq \log L/L_{\odot} \leq
-0.6$. According to BHAC15 this range of $\log L/L_{\odot}$ is roughly
equivalent to $0.8 \geq M/M_{\odot} \geq 0.1$, which spans spectral
types of $\sim$K3 to $\sim$M5 (Kenyon \& Hartmann 1995).  There are a
total 172 targets with measured values of $r\sin i$ with a relative
uncertainty $\le$0.3. A further 22 targets have only upper limits to
their $r\sin i$ and are represented as left censored data
in subsequent analyses.

The peak in the distribution {\it appears} higher than
would be expected for stars with a random alignment of spin axes,
$\overline{r\sin i} \sim \pi/4$, indicating that the stellar radii may
be systematically larger than predicted by the models.

\subsection{Maximum likelihood method}
A maximum likelihood method is used to determine the average radius
ratio or over-radius, $\rho=R/R_m$ of the measured data relative to BHAC15 model radii,
as a function of luminosity. The approach is similar to that used by
Lanzafame et al. (2017) to estimate the over-radius of higher mass
stars in the Pleiades. In the present case the probability of achieving
a particular value of $r\sin i$, written as $\phi(r\sin i|P_j,L_j,\rho)$, is
calculated for individual targets depending on their period, $P_j$,
luminosity, $L_j$, and $\rho$ rather than using uniform probability density
function for all targets with measured $r\sin i$.

\begin{figure}
	\centering
	\includegraphics[width = 80mm]{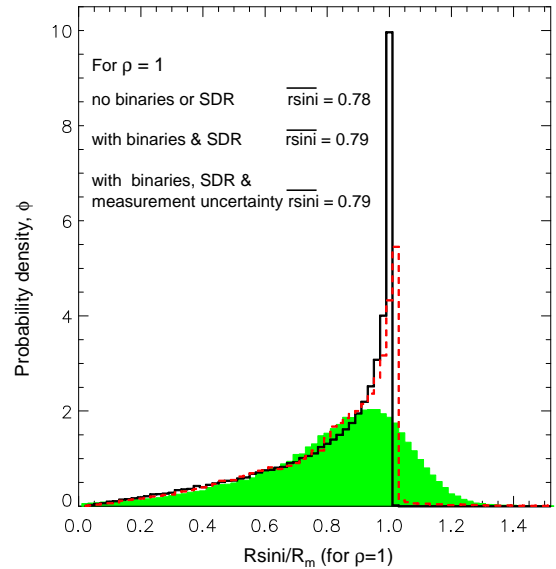}
	\caption{The probability density of measuring a normalised
          radius $r\sin i$ for a representative star of mass,
          $0.4\,M_{\odot}$, SNR$=50$ and $(v\sin i)_p=
          20$\,km\,s$^{-1}$. 
          The black line shows the ideal case of
          precise measurements of $P$ and $v\sin i$ on a single
          star. The dashed red histogram shows the combined effects of
          SDR (assumed to increase $r \sin i$ by a fixed 1 per cent, see
          section 4.2.1) and binarity. The solid green histogram shows the net effect
          of SDR, binarity and measurement uncertainties. }
	\label{fig9}
\end{figure}

\subsection{Probability function for measured data} 
In the ideal case of error-free measurements of $P$ and $v\sin i$ for
single stars, with a random alignment of their spin axes in space, the
probability density increases with inclination as $\phi(i)=\sin
i$. Hence the probability density function of $r\sin i$ is
\begin{equation}
\phi(r\sin i|\rho) =  \rho \tan[\arcsin(r\sin i/\rho)] \; {\rm for} \; r\sin i \le \rho.	
\end{equation}
In practice $\phi$ is modified by the effects of surface differential
rotation, binarity and measurement errors. These effects are investigated
in the following subsections and as an example, 
Figure~10 shows how $\phi$ would be modified for a
representative star of mass, 0.4$M_{\odot}$, SNR$=50$ and $(v \sin
i)_p= 20$\,km\,s$^{-1}$.

\subsubsection{Surface differential rotation}  
 Surface differential rotation (SDR) can lead to
  systematic errors in the estimated radii, because solar-type SDR
  causes the rate of surface rotation to reduce towards the poles
  (Krause \& Raedler 1980). If the starspots responsible for light
  curve modulation are distributed over a range of latitudes then the
  measured rotation rate, $\Omega_m$ will be less than the rotation
  rate at the equator, $\Omega_{\rm eq} $. Reinhold et al. (2013) measured periods for thousands of active stars in the Kepler field. In most cases a second period close to the rotation period was detected which they interpreted as resulting from SDR. 
For low mass stars they found an average difference in rotation rate of
$\Delta \Omega$ = 0.08 radians/day between the two measured periods
that was almost independent of measured period and is a small
fraction of the average angular frequency of our Pleiades targets that
have a measured $r \sin i$, where $\overline{\Omega_m}=$
14\,radians/day. This is in agreement with the analysis of
multi-periodic stars in the Pleiades data by Rebull et al. (2016b),
where no evidence could be found for differential rotation among the
fast-rotating M-dwarfs that are the subject of this paper.
 Taking $\Omega_m-\Omega_{eq}$ as
\textit{approximately} equal to $\Delta \Omega$ then the fractional
increase in measured period compared to the true equatorial period,
$\Delta\Omega/\overline{\Omega_m}$ would be $\leq 1$ per cent. The
corresponding increase in $r\sin i$ will be similarly small. 

The
potential effects of SDR are shown in Fig.~10 for illustrative purposes, but it is
neglected as insignificant in our main analysis and results.

\subsubsection{Binarity} 
A proportion of the targets will be part of unresolved binary systems.
  Short-period binary systems are easily identified from the
  offset in $RV_{rel}$ from the cluster mean and/or double peaks in
  their CCF and these are rejected from the sample. However, a fraction
  of the retained targets will be in longer period spectroscopic
  binaries, resulting in a broadening and shift in $\phi$ for two
  separate reasons, both of which are accounted for with a Monte Carlo
  simulation of the binary population. We assume a binary frequency for
  low-mass Pleiades stars of 30 per cent (Duch{\^e}ne \& Kraus 2013). We
  also adopt the lognormal period distribution and flat mass ratio and
  eccentricity distributions found for field stars by Raghavan et
  al. (2010).

First, the CCF may be broadened due to the unresolved velocity difference
  between the two components. In these cases 
  the measured $v\sin i$ will (on average) be systematically
  larger than the true $v\sin i$ of the primary star, by an amount that
  depends on the
  difference in $RV$ and relative luminosity of the primary and
  secondary. The effect is modelled as described in Appendix A of Jackson et al.
  (2016). For each target, the properties of possible secondary stars
  are drawn at random from the binary distribution described above. 
  The increase in the FWHM of the CCF
  is then estimated as a function of the line of sight velocity of the
  primary and secondary stars (relative to the centre of mass) and the
  relative flux contribution of the secondary at the wavelength of the
  observed spectra. This is done by measuring the FWHM of a Gaussian profile fitted
  to the sum of two separate Gaussian profiles representing the primary
  and secondary stars. The ratio of $v\sin i$ determined from the FWHM
  of the combined profile to the true $v\sin i$ of the primary is averaged to
  determine the bias in $v\sin i$ and hence $r\sin i$ caused by
  binarity. The typical effect is to broaden the distribution of $\phi$
  and produce a small tail of detections with $r\sin i > \rho$. 
  This potential bias in $v\sin i$ decreases with increasing $v\sin i$,
  but the average effect in our sample is to increase the estimated $r
  \sin i$ values by an average of $<2$ per cent.

 Second, the model radius used to calculate $r \sin i$ is systematically
over-estimated in binary systems, because the system luminosity is
larger than the luminosity of the primary star. This leads to an {\it
  under-estimate} of $r \sin i$ if the effect is ignored.
For each binary in the
simulation we estimate the luminosity of the primary and the system and
use this to calculate the bias in $r \sin i$ that is introduced. The
average effect for stars in our sample is to decrease the estimated $r
\sin i$ values by an average of 3 per cent, and the $\phi$ distribution 
is also modified by the appearance of a small ``bump'' at
$r\sin i \sim 0.8$ due to binaries with similar mass components (see Fig.~10).

The net effect of binarity for our sample is to cause a broadening of the
$\phi$ distribution and to decrease the
estimated $r \sin i$ by about 1 per cent, although this
number will depend linearly on the assumed binary frequency and in a
more complex way on the assumed details of the distributions of mass
ratio and orbital period. In what follows we will usually neglect the effects
of binarity, except where the details of the $r \sin i$
distribution become important in section 5.3.2.
	
\subsubsection{Measurement uncertainties} 

Uncertainties in the measurements will broaden the distribution of
  $\phi$ according to the expected fractional uncertainty in $r\sin
  i$. The uncertainty in $P$ is assumed to be small compared to the
  uncertainty in $v\sin i$, hence the fractional uncertainty in $r\sin i$
  also equals $S_{v\sin i}/v\sin i$. A Monte Carlo method is
  used to calculate the effect of measurement uncertainties
  on $\phi$, where the fractional uncertainties in trial values of $r\sin i$
  are the product of the fractional uncertainty in $v\sin i$ given by Eqn.~4 and
  random values drawn from a Student's-t distribution with
  $\nu=3$. Measurement uncertainties broaden the peak in $\phi$ but have
  almost no effect on $\overline{r\sin i}$ (see Fig.~10). 

The effects of
  measurement uncertainties are included in all of our subsequent analyses.

\subsubsection{Multiple periods}

Any uncertainty in $P$ is neglected in our analyses, but in a fraction
of cases -- 36 of the 194 stars used in the final analysis -- Rebull et
al. (2016a) report more than one possible period from the Kepler K2
light curves. In all these cases we have used the first period identified by
Rebull et al. as $P_1$, the most likely rotation period of the star. 
The status and cause of these multiple periods is
discussed in detail by Rebull et al. (2016b). For the fast rotating
M-dwarfs that constitute most of our sample, the multiple periods are
probably due to unresolved binary companions. Given that Rebull et
al. (2016a) find very little correlation between photometric amplitude
and either rotation period or photometric colour for $V-K>2$
(which applies to all our targets) then we expect that in the majority
of these cases the rotation period reported as $P_1$ is the rotation
period of the brighter primary star and that this corresponds to the $v
\sin i$ we have measured. However, there is a possibility that some of
these periods are actually the period of an unresolved, fainter
secondary star and that the $P v \sin i$ value is in error.

Without further information we have no way of knowing the probability
that the measured period, $P_1$, is that of the primary star. To test whether
this could have any implications for our results we simply repeated the
analysis after excluding these stars. The average over-radius (reported
in section 5 and see Table~6) inferred from the filtered dataset was
increased by 1 per cent, which is smaller than the size of the
statistical uncertainties. This small effect is entirely consistent
with our earlier assessment of the effect on the $r \sin i$
determination of including unresolved binary systems if $P_1$ is in fact
the period of the brighter primary in all cases.

\subsection{Censored data}
A probability distribution $\phi(r\sin i|P_j,m_j,\rho)$ is calculated
for the $j^{\rm th}$ target. The value of this function at the measured
$r\sin i_j$ defines the probability assigned to a particular target,
$\phi(r\sin i_j|P_j,m_j,\rho)$. Targets with a relative measurement
uncertainty $>0.3$ are treated as left censored data where an upper
limit $r\sin i_{\rm UL}$ is calculated from the target's period and upper
limit to $v\sin i$.
The probability density for these stars is estimated as;
\begin{equation}
\phi(r\sin i < r\sin i_{\rm UL}|P_i,m_i,\rho) = \frac{\int_{0}^{r\sin i_{\rm
    UL}} \phi(r\sin i|P_i,m_i,\rho)\,d(r\sin i)}{r \sin i_{\rm UL}}\, ,	
\end{equation} 
corresponding to the average value of $\phi$
between 0 and $r\sin i_{\rm UL}$.  

\subsection{Estimating the best fitting over-radius}
The best fit value of $\rho$ is determined by maximising the log-likelihood function; 
\begin{equation}
	\ln \mathscr{L} = \displaystyle\sum_{l}\ln \phi(r\sin i_j) +
         \displaystyle\sum_{m}\ln \phi(r\sin i < r\sin i_{{\rm UL},k})
\end{equation}
where $1 \leq j \leq l$ are the set of targets with measured values of 
$r\sin i$ and $1\leq k \leq m$ are those 
targets with $r\sin i$ upper limits. 
The log-likelihood is computed over a range of values of
$\rho$. The maximum of the likelihood-function, $\ln{\widehat{\mathscr{L}}}$, is
used to define the most likely value of $\rho$, and the standard
deviation of the likelihood-function is used to estimate the uncertainty.

\begin{figure}
	\centering
	\includegraphics[width = 77mm]{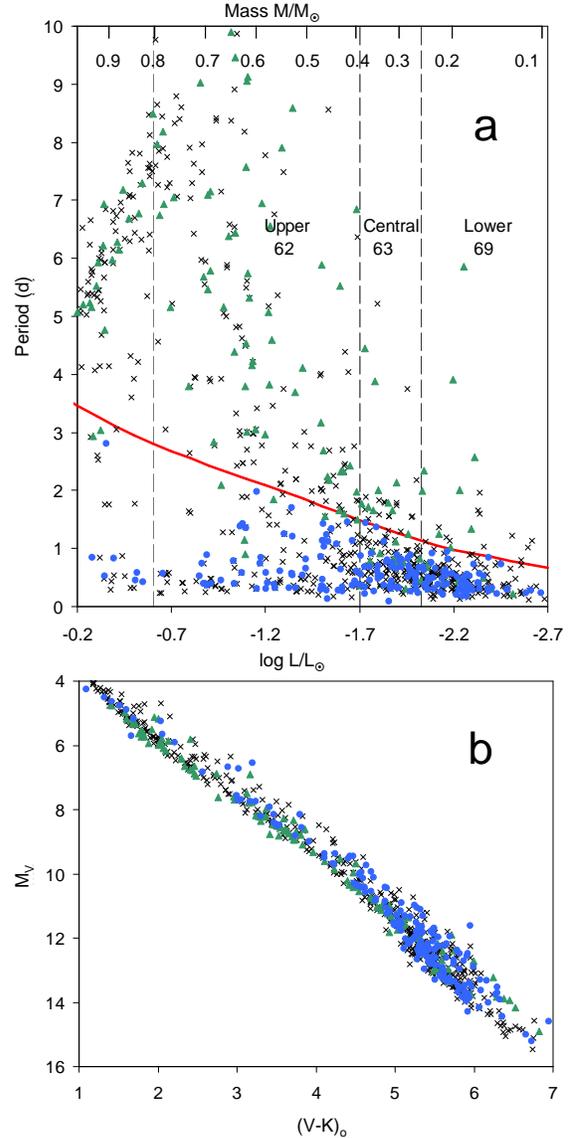}
	\caption{Plot A shows the rotation periods of low mass stars in the Pleiades as
          a function of luminosity.  Circles indicate stars with
          measured spectra; those circles that are filled
          are the subset of stars with a measured $r\sin i$ value with
          uncertainty $\leq 30$ per cent. 
          Crosses show all other stars with measured periods reported by Rebull et al. (2016a), 
          Covey et al. (2016) and Hartman et al. (2010) which were {\it
            not} included in our observations.
                    The solid red line marks the locus of stars with $(v\sin i)_p
          = 10$\,km\,s$^{-1}$; stars below this line are included in
          the $r \sin i$ analysis.
          Dashed vertical lines indicate boundaries between
          the upper, central and lower luminosity bins with numbers
          indicating the number of $r\sin i$ values bin. Plot B shows the $M_V$ versus $(V-K)_0$ colour 
          magnitude diagram using the same symbols and colour coding.}
\label{fig18}
\end{figure}

\section{Results}
In this section we present the measured radii of fast
rotating low mass stars in the Pleiades relative to the radii predicted
for a 120\,Myr cluster using BHAC15 evolutionary models. Figure 11 shows
the measured period versus luminosity of Pleiades targets with $\log L/L_{\odot}
< -0.2$ reported in Rebull et al. (2016a), Covey et al. (2016) and Hartman
et al. (2010), together with those stars for which we obtained
spectroscopy and those stars which were included as part of the $r \sin
i$ analysis. 


\begin{table*}
\caption{The maximum likelihood value of radius ratio, $\rho$, for
  faster rotating low mass stars in the Pleiades relative to the radii
  predicted for a 120\,Myr solar metallicity cluster using the  BHAC15
  evolutionary model. $N_{\rm targ}$ is the number of targets included in
  the calculation of the likelihood function
and $N_{\rm rsin i}$ is the number of those targets with measured values of $r\sin i$.} 
\begin{tabular}{llllllll} \hline
Subset with bins	&	Bin	&	$N_{\rm targ}$	&
$N_{\rm rsini}$	&	$\log L$	&	$\overline{r\sin i}$	&$\ln{\widehat{\mathscr{L}}}$	&	$\rho$			\\\hline
All targets in bin	&	All	&	194	&	172	&	-1.823	&	0.883	&	-75.3	&	1.138	$\pm$	0.013	\\
	&	Lower	&	69	&	63	&	-2.239	&	0.907	&	-24.1	&	1.115	$\pm$	0.028	\\
	&	Central	&	63	&	53	&	-1.874	&	0.863	&	-32.4	&	1.146	$\pm$	0.033	\\
	&	Upper	&	62	&	56	&	-1.308	&	0.876	&	-19.0	&	1.151	$\pm$	0.021	\\\hline
Slower rotators	&	All	&	106	&	88	&	-1.788	&	0.972	&	-25.7	&	1.164	$\pm$	0.024	\\
	&	Lower	&	34	&	30	&	-2.197	&	1.000	&	-5.0	&	1.166	$\pm$	0.043	\\
	&	Central	&	36	&	28	&	-1.874	&	0.991	&	-10.3	&	1.192	$\pm$	0.035	\\
	&	Upper	&	36	&	30	&	-1.315	&	0.925	&	-9.5	&	1.127	$\pm$	0.041	\\\hline
Fast rotators	&	All	&	88	&	84	&	-1.866	&	0.790	&	-49.2	&	1.121	$\pm$	0.020	\\
	&	Lower	&	35	&	33	&	-2.281	&	0.820	&	-17.4	&	1.087	$\pm$	0.029	\\
	&	Central	&	27	&	25	&	-1.875	&	0.720	&	-19.4	&	1.081	$\pm$	0.039	\\
	&	Upper	&	26	&	26	&	-1.297	&	0.821	&	-9.1	&	1.166	$\pm$	0.034	\\\hline
Lower amplitude	&	All	&	98	&	86	&	-1.886	&	0.803	&	-56.0	&	1.056	$\pm$	0.022	\\
light curves	&	Lower	&	37	&	33	&	-2.241	&	0.833	&	-16.5	&	1.063	$\pm$	0.031	\\
	&	Central	&	33	&	28	&	-1.886	&	0.751	&	-25.1	&	1.057	$\pm$	0.046	\\
	&	Upper	&	28	&	25	&	-1.415	&	0.825	&	-13.5	&	1.080	$\pm$	0.050	\\\hline
Higher amplitude	&	All	&	91	&	83	&	-1.795	&	0.965	&	-9.2	&	1.165	$\pm$	0.017	\\
light curves	&	Lower	&	32	&	30	&	-2.237	&	0.987	&	-4.9	&	1.167	$\pm$	0.038	\\
	&	Central	&	29	&	24	&	-1.866	&	0.994	&	-4.0	&	1.181	$\pm$	0.037	\\
	&	Upper	&	30	&	29	&	-1.255	&	0.920	&	-0.1	&	1.165	$\pm$	0.025	\\\hline
\end{tabular}
\label{reference}
\end{table*}

\begin{table*}
\caption{Sensitivity of the inferred values of the
  over-radius $\rho$ (overall and in the three luminosity bins introduced in
  section 5.1), to the assumed age and distance of the cluster and to
  parameters set in the data reduction pipeline. $N_{\rm targ}$ and
  $N_{\rm rsin i}$ are as defined in Table~5. } 
\begin{tabular}{llcccccr} \hline
	&	&N$_{targ}$/ 	& \multicolumn{4}{c}{$\rho$ for bin:} &		$\Delta \rho^{*}$	\\
	&	&N$_{r\sin i}$ &		Upper	&	Central
  & Lower		& All	& \\\hline 		
Case 0 - &  Reference: age 120\,Myr, distance 136.2\,pc	&	194/172	&	1.15	$\pm$	0.02	&	1.15	$\pm$	0.03	&	1.12	$\pm$	0.03	&	1.138	$\pm$	0.013	&	---	\\
Case 1 - &  Model age reduced to 80Myr	&	197/173	&	1.13	$\pm$	0.02	&	1.11	$\pm$	0.03	&	1.08	$\pm$	0.03	&	1.109	$\pm$	0.018	&	-0.029	\\
Case 2 - &  Model age increased to 160Myr	&	194/172	&	1.15	$\pm$	0.02	&	1.16	$\pm$	0.03	&	1.13	$\pm$	0.03	&	1.141	$\pm$	0.016	&	+0.003	\\
Case 3 - & Increased distance distance 140.0\,pc	&	196/173	&	1.15	$\pm$	0.02	&	1.11	$\pm$	0.03	&	1.08	$\pm$	0.03	&	1.120	$\pm$	0.015	&	-0.018	\\
Case 4 - &  Reduced distance distance 132.4\,pc	&	192/172	&	1.17	$\pm$	0.02	&	1.17	$\pm$	0.03	&	1.13	$\pm$	0.03	&	1.156	$\pm$	0.015	&	+0.018	\\
Case 5 - &  Compensation for binarity and SDR	&	194/172	&	1.16	$\pm$	0.03	&	1.17	$\pm$	0.03	&	1.14	$\pm$	0.03	&	1.152	$\pm$	0.016	&	+0.014	\\
Case 6 -&Minimum $(v\sin i)_p=$ 15\,km\,s$^{-1}$	&	167/153	&	1.15	$\pm$	0.02	&	1.15$ \pm$	0.03	&	1.11	$\pm$	0.03	&	1.134	$\pm$	0.014	&	-0.004\\
Case 7 -& Excluding targets with multiple periods	&	158/138	&	1.17	$\pm$	0.03	&	1.17	$\pm$	0.03	&	1.12	$\pm$	0.03	&	1.150	$\pm$	0.014	&	+0.012	\\
Case 8 - &  Increase value  of FWHM$_0^{**}$	&	194/166	&	1.14	$\pm$	0.02	&	1.12	$\pm$	0.03	&	1.10	$\pm$	0.03	&	1.124	$\pm$	0.015	&	-0.014	\\
           &  (i) slower rotators	&	106/83	&	1.11	$\pm$	0.04	&	1.17	$\pm$	0.04	&	1.14	$\pm$	0.05	&	1.137	$\pm$	0.024	&	-0.027	\\
          &   (ii) Faster rotators	&	88/83	&	1.16	$\pm$	0.04	&	1.08	$\pm$	0.04	&	1.08	$\pm$	0.03	&	1.117	$\pm$	0.021	&	-0.004	\\
Case 9 - &  Upper 90\% of light curve amplitudes	&	173/158	&	1.16	$\pm$	0.02	&	1.16$ \pm$	0.03	&	1.13	$\pm$	0.03	&	1.148	$\pm$	0.016	&	+0.010	\\
\hline
													
	\multicolumn{8}{l}{* Change in $\rho$ calculated for data over the full luminosity range relative to the reference case.}\\ 
	\multicolumn{8}{l}{** The value of FWHM assumed for slow-rotating stars in the calculation of $v\sin i$ is increased by 0.5\,km\,s$^{-1}$ ($\sim$3 times uncertainty in FWHM$_0$)}\\ 

\end{tabular}
\label{sensitivity}
\end{table*}

\subsection{The estimated over-radius}

The data were allocated to three luminosity bins for analysis with roughly equal
numbers of targets per bin (see Fig.~11). The upper bin spans a
relatively wide range of mass (0.4 to 0.8\,M$_{\odot}$, estimated
from the BHAC15 models) and includes both fast rotating stars and stars 
that appear to be in transition between the gyrochronological ``C sequence'' and
slower ``I sequence'' defined for F-K stars by Barnes (2003, 2007). 
The central and lower mass bins are more densely
populated, but consist of almost exclusively of fast rotating stars. 
It is clear that the stars included in the
$r \sin i$ analysis represent a subset of the total population that is heavily
biased towards faster rotators with $P<2$ days and the majority with
$P<1$ day. It is the radius of these faster rotating stars
that is reported here.

The results of the maximum likelihood analysis are presented in
Figs. 8 and 12 and summarised in Table 5. Figure~12 shows the
main result of this paper: the targets considered here
have an average over-radius $\rho =1.138 \pm 0.013$ relative to the
radius-luminosity relation predicted by the solar-metallicity 
evolutionary models of BHAC15 at an age of 120\,Myr. Also shown in Fig.~12 are
over-radii for the upper, central and lower mass/luminosity bins, where the
maximum likelihood analysis has been conducted separately for each
bin. The over-radius in each bin is significantly
larger than unity, and consistent with
the mean over-radius. There is no strong evidence for any variation 
in the level of over-radius across this luminosity and mass range.

The results in Table 5 assume an age for the Pleiades of 120\,Myr,
a distance of 136.2\,pc and neglect the small effects of surface
differential rotation and binarity discussed in section 4.2.  Table~6
shows the effects of changing the assumed age and distance or including the
effects of SDR and binarity on the calculated over-radius. 
Varying the assumed age between 80 and 160\,Myr gives $1.11 <
\rho < 1.14$ (the change in $\rho$ is asymmetric because the stars are
already close to the ZAMS at 120 Myr). Altering the distance between
132.6\,pc and 140.0\,pc (corresponding to the 3$\sigma$ limits reported
by Melis et al. 2014) changes the derived luminosities and hence the
predicted radii and gives $1.12< \rho <1.16$. The combined effects of
SDR and binarity are small, act in opposite directions and depend to
some extent on the assumed binary properties of the sample and latitude
distribution of spots. The net effect of including these would be only
a 1 per cent increase in $\rho$, but given the uncertainties we do not
include this in our final estimate. Table 6 also shows the effects of 
excluding stars with multiple periods and of changing the minimum
$(v \sin i)_p$ threshold for stars included in the analysis (see section
4). These also lead to only $\sim 1$ per cent changes in the main results.

Combining expected levels of
uncertainties in age and distance with the precision-based uncertainty
shown in Table 5 gives a final average over-radius $\rho=1.14\pm0.02$
relative to the solar metallicity BHAC15 model where the uncertainty
represents the 68 per cent confidence intervals for an age of $120 \pm
20$\,Myr and a distance of $136 \pm 2$\,pc.

\begin{figure}
	\centering
	\includegraphics[width = 80mm]{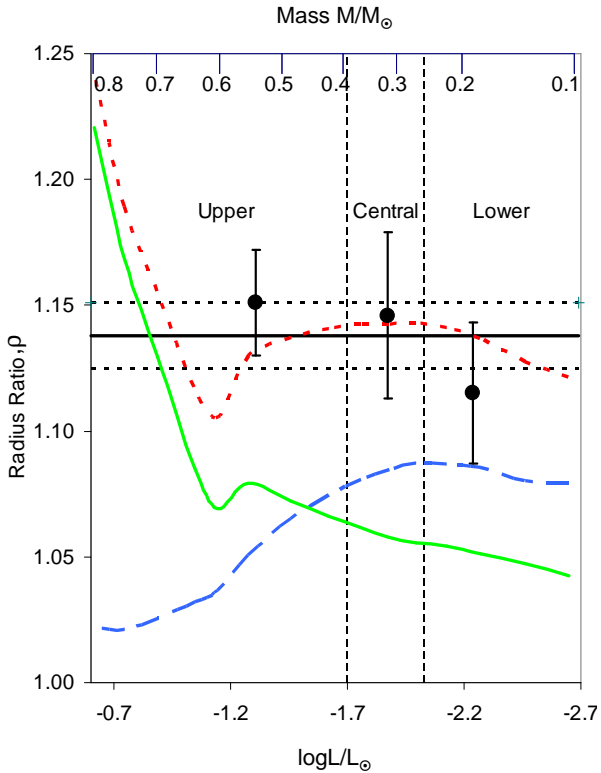}
	\caption{
The estimated over-radius for the subsample of Pleiades stars included
in the maximum likelihood analysis described in section 4. The
over-radius is with respect to the radius predicted by the 
BHAC15 models for 120\,Myr solar
metallicity stars at a given luminosity. The horizontal line shows the mean over-radius,
for all the data considered, with dashed lines indicating the 1-sigma
confidence interval. The individual points with error bars
show the estimated mean over-radius and corresponding uncertainties for
stars in three luminosity/mass bins. The mass scale at the top of the
plot is based on the same BHAC15 model. A green
solid line shows the predicted effect of radius inflation due to
magnetic inhibition of convection (Feiden et al. 2015); the blue
dashed line shows the predicted effect of starpots with an
effective dark spot coverage of $\beta =0.16$ (see section 6.1). The
red dotted line shows the combined effect of both magnetic inhibition of
convection and starspots with $\beta =0.16$.}
\label{fig11}
\end{figure}

\subsection{Comparison of fast and slow rotators and with interferometric radii}
In Fig.~13 the over-radius is shown in the three luminosity/mass bins, but
now also dividing the stars into faster- and slower-rotating subsamples
(note that all these stars should be considered fast rotators when compared
with the parent sample of Pleiades stars and that we
cannot resolve projected radii for slowly rotating stars in the Pleiades).
For the upper and central luminosity bins, the faster rotators are defined as
those having with $P<0.55$\,d. For the lower bin the split is made at
$P<0.4$\,d and this gives roughly equal numbers of stars in each
subsample. Both the central
and lower mass bins show a marginal ($\sim 2 \sigma$) difference in over-radius between the
faster and slower rotators, with the {\it slower}
rotators showing the larger over-radius. No significant difference is seen
for stars in the upper mass bin. 

Figure~13 also shows $\rho$ for individual field stars
based on interferometric measurements of stellar radii report by
Boyajian et al. (2012) plotted against luminosity values derived from their
2MASS $K$ magnitudes and $V-K$ colours in the same way as the Pleiades
data. For these stars, the measured radii are compared to a 5\,Gyr solar
metallicity BHAC15 isochrone, although neither the age or
metallicity are well constrained for most of these targets. There are 19 stars with
reported radii in the upper bin, with a weighted mean
$\rho = 1.026 \pm 0.007$. The two lower bins contain only 4 stars
with significantly scattered normalised radii, so no useful
comparison can be made with the models. There is thus evidence for a
small over-radius {\it with respect to the solar metallicity models}
for the field M-dwarfs, but it is much smaller than the over-radius
found in the Pleiades.

\begin{figure}
	\centering
	\includegraphics[width = 75mm]{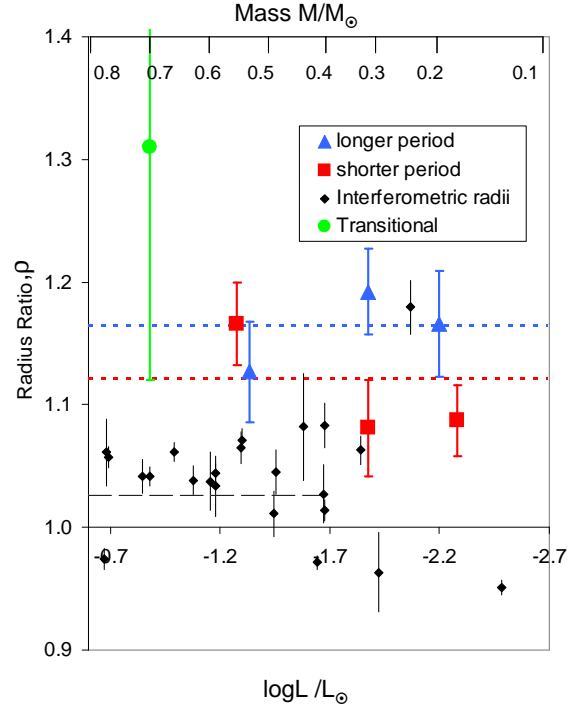}
	\caption{
The influence of rotation rate on the over-radius of Pleiades
stars relative to the predictions of a BHAC15 120\,Myr isochrone. Pleiades targets 
in each luminosity bin are split by period
into faster and slower rotating subsamples (see section 4.5). Also
shown are over-radii for low-mass field stars, derived from
interferometric radius measurements in Boyajian et al. (2012), and with
respect to a 5 Gyr solar-metallicity isochrone from BHAC15.
The green point shows a previously reported measurement of $\rho$ for
Pleiades stars of $0.6<M/M_{\odot}<0.8$ and periods in the transition
between the C and I gyrochronological sequences defined by Barnes
(2007) ($P>2$d, see Table~2 of Lanzafame et
al. 2017).}
\label{fig12}
\end{figure}

\begin{figure}
	\centering
	\includegraphics[width = 75mm]{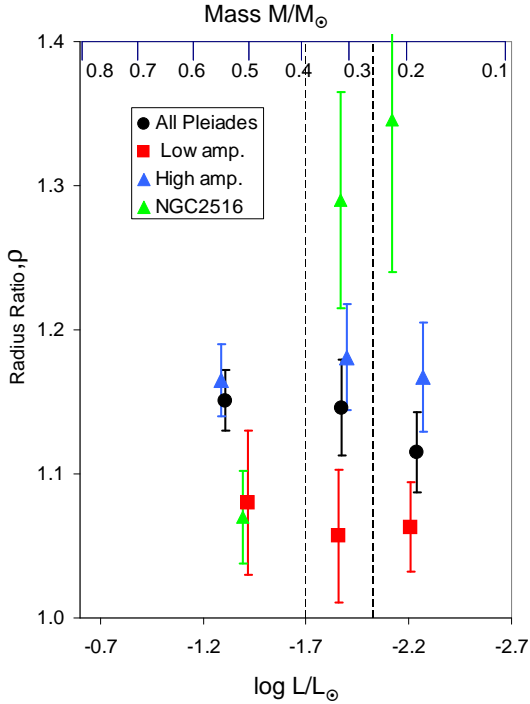}
	\caption{
The influence of light curve amplitude on the estimated over-radius. 
Pleiades targets in each luminosity bin are split
into subsamples with higher and lower spot-induced light curve
amplitudes (see section 5.3.1). Green triangles show the over-radius of targets in 
NGC\,2516 estimated using previously
reported measurements of $P$ and $v\sin i$ (see section 6.2 and Jackson et al. 2010a).}
\label{fig13}
\end{figure}

\subsection{Biases and selection effects due to 
spot coverage and the $\sin i$ distribution.}

The analysis described above explicitly assumes that the spin axes of
targets are randomly distributed in space giving a probability density
$\phi(i)=\sin i$. There are a number of reasons why this may not be
true: it is easier to resolve $v\sin i$ in targets with higher
values of $\sin i$; measurements of period may be biased towards
targets with larger $\sin i$ since these would exhibit larger light
curve amplitudes due to starspot modulation; or the spin axes may be
partially aligned, 
yielding a bias that could result in either a higher or lower mean $\sin
i$ and perhaps a narrowing of the $\sin i$ distribution. 

The possibility of bias in the measured over-radius due to the
inability to measure $v\sin i$ at low inclinations is already circumvented in
the present analysis by explicitly including targets with upper limits in
$v\sin i$ as left-censored data. The remaining sources of bias are
considered separately below.

\subsubsection{Selection of stars with higher amplitude light curves}
The possibility of bias due to selection effects in the period
measurements depends on the completeness of the period data;
i.e. whether periods are available for a representative sample of
stars, including those with low inclinations.  Figure 14 illustrates
the potential effect of incomplete sampling of period data on the
measured over-radius -- selecting stars with higher spot-modulated
light curve amplitude is expected to preferentially select stars with
higher $\sin i$.  Selecting for analysis only the 50 per cent of
Pleiades targets with the highest light curve amplitude (taken from
Rebull et al. 2016a) increases the estimated value of $\rho$ by $\sim 3$
per cent compared to an analysis of the entire sample. This is
actually less than would be expected if light curve amplitude depended
solely on $\sin i$, since $\overline{\sin i}$ increases from 0.785 for
a randomly distributed sample to 0.96 for a group of objects with $\sin
i$ restricted to be above the median of a random distribution.

We do not believe our over-radius results for the Pleiades can be
biased by anything like this amount. Most of the targets (189 out of 192) have
periods measured from K2 light curves. Rebull et al. (2016a)
measured periods for 92~per cent of the Pleiades targets
observed. For half of the remaining stars periods were not measured 
because of non-astrophysical or instrumental effects, leaving just 4
per cent without measured periods that could be Pleiades members with
low inclination angles or targets with very long periods or very few
(or very symmetrical) star spots.

The effect of missing a small proportion of targets with the lowest
amplitude light curves can be assessed by evaluating $\rho$ using only
those targets from our list with the top 90 per cent of light curve
amplitudes. This produces a small (1 per cent) increase in $\rho$
(see Table~6), indicating that our sample of measured periods, which is
complete to $\sim$96 per cent provides an
almost unbiased estimate of $\rho$.

\begin{figure}
\	\centering
	\includegraphics[width = 85mm]{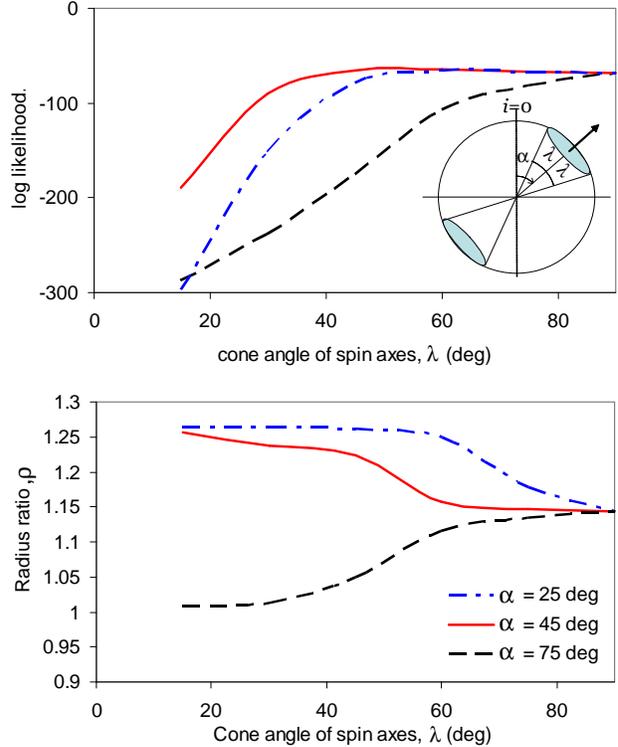}
	\caption{The effect of partial alignment of stellar spin
          axes. The upper plot shows the variation of
          $\ln{\widehat{\mathscr{L}}}$ with cone angle $\lambda$, for
          3 different values of cone inclination $\alpha$, relative to the line of sight. The lower plot shows the variation of $\rho$ over the same parameter range}
\label{fig14}
\end{figure}

\subsubsection{Alignment of stellar spin axes} 
Jackson and Jeffries (2010a) investigated the effects of partial
alignment of spin vectors
by modelling cases where spin axes are uniformly distributed
inside a cone and zero elsewhere. The cone semi-opening angle, 
 $\lambda$, determines the degree of alignment,
and the mean inclination of the stars within the cone is $\alpha$ (see Fig.~15).
In this case the probability function $\phi(\sin i|\rho)$ in eqn.~5 is replaced by a more 
complex function, $\phi(\sin i|\rho,\alpha,\lambda)$  calculated using
a Monte Carlo method  (see eqns.~2 to 6 in Jackson \& Jeffries 2010a).

Figure~15 shows the effect of partial alignment of
stellar spin axes on the maximum log-likelihood (see Eqn. 7) and the
derived value of $\rho$ for
$15^{\circ} < \lambda < 90^{\circ}$ (the upper limit
corresponds to random alignment of the spin axes). In this analysis the
effects of SDR and binarity {\it are} included because whilst they have
little effect on the mean inferred $r \sin i$, they {\it do} have a
small, but non-negligible effect on the detailed shape of $\phi$. Results
are shown for three values of $\alpha$:
\begin{itemize}
	\item For $\alpha=25^{\circ}$ the spin axes are
          aligned close to the line of sight, such that the average
          value of $\sin i$ is lower than the case of a uniform
          distribution. Consequently a higher value of $\rho$ is
          required to match the observed set of $r\sin i$ values.
	
        \item
	For $\alpha=45^{\circ}$ the spin axes are aligned as shown in
        the sketch in the upper panel of Fig.~15. If $\lambda
        <45^{\circ}$ then $\overline{\sin i}$ is lower than the uniform
        case and hence $\rho$ is higher. At larger $\lambda$ values
        both the maximum likelihood and $\rho$ are similar to the case of a uniform distribution.
	
	\item For $\alpha=75^{\circ}$ and small $\lambda$ the spin axes are
          aligned almost perpendicularly the line of sight. This both
          increases $\overline{\sin i}$ and suppresses the expected
          number of targets with low $r\sin i$ (relative to the mean)
          and therefore 
          provides a poor match to the measured distribution of $r\sin
          i$. This allows us to say that if $\alpha$ is as large as
          this, then $\lambda >80^{\circ}$ degrees.
	
\end{itemize}

\section{Discussion}

\subsection{The over-radius in low-mass Pleiades stars}
Observations of low-mass, short-period eclipsing binaries reveal
that their components may be inflated by $\sim$10 per cent at a given mass
compared with the usual evolutionary models. We have found a similar
phenomenon here. The average over-radius in our sample of
fast-rotating, low-mass Pleiades stars is
$14\pm2$ per cent {\it at a given
luminosity}, which according to the polytropic
models discussed by Jackson \&
Jeffries (2014a), is equivalent to a $\sim$9 per cent
over-radius {\it at a given mass}. 

There is no evidence for any mass or luminosity dependence of this
over-radius across the range covered by our sample. In particular, we
have no evidence that the inflation changes markedly as we move from stars with
higher luminosities that have radiative cores, to lower luminosity stars
that should be fully convective. For non-inflated stars aged 120\,Myr
the BHAC15 model shows a radiative core developing at the transition between the central and lower bins in Fig.~12 ($M>0.3M_{\odot}$, $\log L/L_{\odot}>-2.0$).

It should be stressed that the inferred over-radius is with respect to
the evolutionary models of BHAC15 (although the comparison with the
models of Dotter et al. 2008 is almost identical). The evolutionary
models might fail to correctly predict the measured radii for a number
of reasons, although uncertainties in the assumed age and distance are
already incorporated into the error bars on the results.
Before concluding that the over-radius is due to magnetic activity, as
opposed to some other deficiency in the models, we
should compare the same models to the measured radii of older, less
magnetically active, but otherwise similar stars.
The Boyajian et al. (2012) sample of stars with interferometric
measurements of angular radii offers this test for the higher mass
stars ($>$0.4\,$M_{\odot}$) in our sample (see Fig.~13). Stars in this upper
mass bin have a weighted mean over-radius of $2.6 \pm 0.7$ per cent relative to
a 5\,Gyr solar metallicity isochrone.  Hence the over-radius of the
higher mass Pleiades stars relative to the measured radii of inactive
field stars of similar luminosity is $\sim 10$ per cent, although a
detailed comparison is hampered by
uncertainties in the age and metallicities of the field stars. 

Radius inflation at a given luminosity leads to lower
effective temperatures and lower core temperatures in contracting PMS
stars. Work by Jackson \&
Jeffries (2014a,b); Somers \& Pinsonneault (2015a,b) and Feiden (2016) has considered
how this influences the determination of ages and masses of PMS stars
in the Hertzsprung-Russell diagram and the onset and rate of lithium
depletion in their photospheres as it is burned in the core. The amount
of radius inflation we have determined is consistent with what was
assumed or modelled in these works and so the consequences will also be
similar. 

In a cluster like the Pleiades, ages come from either the main-sequence
turn-off or the ``lithium depletion boundary'' (LDB) -- the luminosity
below which Li is preserved in the interior of a fully convective
low-mass PMS star (e.g. Stauffer et al. 1998; Jeffries \& Oliveira
2005). The former is unaffected by radius inflation in low-mass stars,
but the latter may be. We caution the reader that the LDB in the
Pleiades occurs in objects close to the substellar boundary at $\log
L/L_{\odot} \simeq -2.9$ and radius inflation has not yet been
established at these low-masses. However, if stars near the LDB were
inflated by 14 per cent then the calculations presented by Jackson \&
Jeffries (2014b; calculated for inflation due to spots, but valid for
inflation by any other cause) suggest the LDB age should be increased
by 11 per cent, from 125\,Myr to 139\,Myr.

Somers \& Pinsonneault (2015a) and Somers \& Stassun (2017) have
suggested that inflation varies between roughly zero for the slowest
rotators and 15 per cent for the fastest rotators, and could
explain the observed rotation-dependent Li depletion pattern in
Pleiades K-dwarfs.  These stars are at the upper end of the mass range
consider here, but the overall level of radius inflation we measure
in the fastest rotating cluster members, is in agreement with
this hypotheses.

The effects of radius inflation are likely to be even more significant
if present at at younger ages. Jeffries et al. (2017) showed, using the
example of the Gamma Velorum cluster, that ages inferred from the
Hertzsprung-Russell diagram could be doubled by 10 per cent inflation
at a given luminosity (slightly less than found here) and that inferred
masses would also be significantly underestimated by non-magnetic
models, particularly at the lowest stellar masses.

\subsubsection{The possible causes of radius inflation}

That an over-radius has been observed in the Pleiades whilst the models
work reasonably well for older fields stars is circumstantial evidence
that magnetic activity and rotation are the factors responsible for the
over-radius; although some other age-dependent variation in the
physical model could conceivably lead to the observed results.

There are two main "flavours" of magnetic model that might provide an
explanation for the observed over-radii - the magnetic inhibition of
convection at and just below the surface in layers with significant
super-adiabaticity (Feiden \& Chaboyer 2012, 2014), or the blocking of
radiative flux from the surface of the star by cool, magnetic starspots
(Jackson \& Jeffries 2014a; Somers \& Pinsonneault 2015b). These models
predict a different behavior of over-radius as a function of
luminosity. 

Magnetic inhibition becomes less effective as the convection zone
deepens and the stars become fully convective (Feiden et al. 2014).
The solid line in Fig.~12 shows the over-radius using the Dartmouth
code modified for the effects of magnetic field (Feiden, Jones \&
Chaboyer 2015) relative to the "standard" Dartmouth model (Dotter et
al. 2008), assuming a surface field strength of 2.5\,kG as described by
Malo et al. (2014).  Our results suggests that the magnetic inhibition
models, based on an approximate equipartition magnetic field strength
at the stellar surface do not inflate the low luminosity stars in our
sample sufficiently (by a factor of two).

Conversely, the effect of a given coverage of starspots becomes larger
in fully convective stars (Spruit \& Weiss 1986).  Fang et al. (2016)
used the TiO band strengths measured in LAMOST spectra to estimate the
spot coverage and temperatures of low mass stars in the Pleiades.
Their results can be used to estimate an ``effective spot coverage'',
$\beta$, defined as the fraction of stellar flux blocked by starspots
compared to the flux of an immaculate photosphere (equivalent to
$f{_s}^{\prime}$ in Fig.~11 of Fang et al.). Comparing target lists we
find 22 stars analysed by Fang. et al. with a measured $r\sin i$ in our
analysis.  The average value of $\beta$~is 0.16, with a
dispersion of 0.09. This can be used to model the effects of spot
coverage on the \textit{average} stellar radii (Spruit 1982). The
dashed line in Fig.~12 shows the predicted radius ratio for $\beta
=0.16$ as a function of $\log L$, estimated from a linear interpolation
of the calculations of Somers \& Pinsonneault (2015a), that use a
version of the YREC evolutionary code (van Saders \& Pinsonneault 2012)
modified to include starspots.  If radius inflation were caused {\it
  solely} by starspots then this would require $\beta \simeq 0.3$ for
the higher mass stars in our sample, decreasing to $\beta=0.2$ at lower
masses where the effects of a given spot coverage are stronger.

These spot coverages are only a little larger than suggested by Fang et
al. (2016) but it is possible that $\beta$ has been underestimated by
their simple two-component modelling of the optical spectra.
Alternatively, one could have both mechanisms in operation, with the
more modest \textit{average} spot coverage measured by Fang et al. (2016)
($\beta=0.16$) plus magnetic inhibition of convection by an
equipartition surface magnetic field ($\sim 2.5$\,kG), and the sum of
these two would match the measured over-radii reasonably well (dotted line in Fig.~12).

\subsubsection{Influence of rotation rate}
Given that we are hypothesising that strong, dynamo-induced, magnetic
fields are the root cause of the over-radius, it is interesting to investigate
whether there is any dependence on rotation rate.  Lanzafame
et al. (2017) found a complex behavior in Pleiades K-stars and
suggested, albeit with low number statistics, that stars with
intermediate rotation rates (those between the C- and I-sequences
described by Barnes) had larger over-radii than stars with the
fastest rotation rates. By splitting our sample into 
fast and slow(er) rotating halves we have found marginal evidence (see Fig.~13) that 
partially supports Lanzafame et al.'s result -- though we note (i) that
our sample does not contain many stars rotating as slowly as those
included in Lanzafame et al.'s sample and (ii) that there is no
suggestion of separate C- and I-sequences in the
rotation period data of lower mass stars ($M<0.6M_{\odot}$) in the
Pleiades (see Fig.~11). The slow(er) sample has a mean over-radius  
about 2-sigma higher than the fastest rotators, though note that all of
these stars rotate fast enough to be considered magnetically saturated.

It is possible that this difference is linked to the structure of the
star and possibly the presence of a radiative core. When considered in
three luminosity bins (Fig.~13), our results suggest that
any difference in over-radius is confined only to the lowest luminosity stars
and in fact there is no significant difference for the high luminosity end of our
sample where there is overlap with the sample considered by Lanzafame et
al. (2016). 

We would caution against ascribing too much significance to this result
at this stage, since the samples may be affected by analysis biases that
could separate the over-radii of fast- and slow(er)-rotators. 
For example we are not able to measure $v\sin i$ on slowly
rotating targets. Whilst we have taken steps to address this bias in
our analysis it is possible that some uncertainties remain.  There is
also the possibility of uncertainty in the zero-point of the $v\sin i$
calibration. As pointed out by Hartman et al. (2010), if the zero-point
is too low then this could result in a significant over-estimate of
$v\sin i$ (and hence $r \sin i$) for stars with the smallest resolvable
$v \sin i$, but much less effect for the fastest rotators. To test
this, we artificially raised the zeropoint by 0.5\,km\,s$^{-1}$, which
is far beyond any likely statistical error in our zero-point (see Table~6). This
reduces the overall level of inflation by 1 per cent for the entire
sample whilst decreasing the "gap" between faster and slower rotators
  by 2 per cent. 
  
A more intriguing possibility is that this difference is real. Reiners
\& Mohanty (2012) have claimed that the angular momentum loss rate due
to a magnetically coupled wind is much more strongly dependent on the
stellar radius ($\propto R^{16/3}$) than assumed in previous work
(e.g. Kawaler et al. 1988) and more strongly than it depends on
rotation rate ($\propto \Omega$ in the magnetically saturated regime,
which all our stars are). From this perspective, two similar stars with
radii that differ by 10--20 per cent would have quite different angular
momentum loss rates. Even if greater radius inflation were initially
caused by more rapid rotation and greater magnetic activity, the
consequent spin-down timescale could be much shorter than the thermal
timescale on which an inflated star could react to a slower rotation rate 
and so we might expect to see that the stars that have begun
to spin down are indeed those with larger radii. A more detailed
analysis of this possibility is beyond the scope of this paper and
perhaps not yet warranted by the quality of the data.

\subsection{The discrepancy with NGC 2516}
In Jackson et al. (2009) we undertook a similar analysis of spectra for
low-mass stars with known rotation period in NGC 2516 - a cluster with
a similar age and metallicity to the Pleiades. The results 
differed in that the deduced over-radius at the lowest
masses considered in that paper ($\simeq 0.25M_{\odot}$) reached
$\sim 40$ per cent. Stars with higher masses were in reasonable
agreement with what we find for similar stars in the Pleiades.

Here, we have adopted a maximum likelihood technique including stars
with upper limits in $v\sin i$ as left censored data. In the NGC\,2516
work, we gave equal weight to each measured $r\sin i$ value and allowed
for left-censored data 
by adopting a lower cut-off in $\sin i$, below which
$r\sin i$ could not be measured. Re-running the NGC 2516 dataset
through the current analysis pipeline (and also using the BCAH15 models as
our baseline), we instead find $\rho =1.31 \pm 0.06$ for data in the two lower
luminosity bins (see Fig.~14). This value of $\rho$ is still significantly higher than
the average over-radius measured from Pleiades data.

A substantial and pertinent 
difference between the Pleiades and NGC\,2516 datasets is the fraction
of observed targets with measured rotation periods. Jackson and Jeffries (2012)
reported that less than half of the NGC\,2516 members monitored (from
the ground) by Irwin et
al. (2007) had subsequently derived rotation periods. This fraction was
about 50 per cent in the higher mass bins, dropping to
30 per cent for the lowest luminosity bin. Selecting a similar subset of
Pleiades targets, (those with the top 40 per cent of light curve
amplitudes in Rebull et al. 2016a) yields $\rho =1.18 \pm 0.04$ for
stars with $M/M_{\odot} < 0.40$. Whilst this is 4 per cent higher than 
obtained using the full range of amplitudes it is still $13 \pm 7$ per cent
lower than found for similar NGC\,2516 targets. We have
been unable to identify any other significant systematic differences between the
two data sets that might account for this remaining discrepancy, if indeed it is real.

This comparison has highlighted the importance of having as
near complete a set of period data as possible when estimating
$\rho$. Whilst the maximum likelihood method used here includes targets
with low $v\sin i$ as left censored data it neglects targets without
measured periods and this can lead to a bias in the $\sin i$ distribution. Details of
cluster members {\it without} measured periods are often not reported
in catalogues of rotation periods and this fraction can be high,
especially for ground-based surveys with limited sensitivity to
low-amplitude modulation.

\subsection{The $\sin i$ distribution}

Corsaro et al. (2017) used
asteroseismology-based estimates of inclination angles to claim that
the distribution of spin-axis vectors is not random among stars with
$M>M_{\odot}$ in two old open clusters in the main Kepler
field. They attribute the strong alignment effect they find to the
formation of these clusters from a collapsing cloud with a high ratio
of rotational to turbulent kinetic energy and the inheritance of some
of this angular momentum by the forming stars, especially those with
higher masses. From their simulations, Corsaro et al. suggest that this
effect may be much weaker in the lower mass stars ($<0.7 M_{\odot}$)
that constitute most of our Pleiades sample.  If we assume that the dispersion
in $r\sin i$ that we see is mostly caused by a variation in $\sin i$
and not by a star-to-star variation in over-radius, then our
observations put constraints on how narrow the distribution of
$\sin i$ could be.

Figure 15 showed the effects of alignment of spin axes on
$\ln{\widehat{\mathscr{L}}}$ and $\rho$ when spin axes are uniformly
distributed over a cone with half opening angle $\lambda$ and
average inclination $\alpha$ relative to the line of sight (see section
5.3.2). There is no strong evidence of preferential alignment. A model with
$\lambda=90^{\circ}$, equivalent to a random distribution, has $\ln{\widehat{\mathscr{L}}}$ that is not significantly lower than the best fitting model with $\lambda<90^{\circ}$ according to the Bayesian information criterion (BIC; $\Delta {\rm BIC} \sim4$). The minimum cone angle that provides a similar value of
$\ln{\widehat{\mathscr{L}}}$ to a random distribution of spin axes is
$\lambda \geq 30^{\circ}$ if $\alpha <45^{\circ}$ i.e. a strongly
aligned spin axis distribution with
$\lambda <30^{\circ}$ does not match the measured distribution of
$r\sin i$ for any mean inclination. If $\alpha >45^{\circ}$, which is
$>70$ per cent likely for a randomly distributed $\alpha$, then the
lower limit to $\lambda$ becomes much larger.

Although the measured distribution of $r\sin i$ could be matched by a partial
alignment of spin axes with $\lambda \geq 30^{\circ}$, the
most likely value of $\rho$ in those cases is always similar to, or larger than,
the value obtained by assuming a random distribution of spin axes (see Fig.~15). Thus
the mean over-radius of 14 per cent shown in Fig.~12 and Table~5 is
the {\it minimum} that provides an acceptable fit to the measured
data. 

\section{Summary}

Precise measurements of rotation periods from the Kepler K2 survey of
the Pleiades have been combined with new, precise measurements of
rotational broadening for the same stars, in order to estimate their
projected radii. Using a maximum likelihood analysis technique and
assuming random spin-axis orientation, the average radius of
fast-rotating ($P \sim 2$ days or less), low-mass ($0.1 \leq
M/M_{\odot} \leq 0.8$) Pleiades members is $14\pm2$ per cent larger
than predicted for stars of the same luminosity by the
solar-metallicity Baraffe et al. (2015) and Dotter et al. (2008) models
for an assumed age of 120\,Myr. The analysis considered unresolved
binarity, differential rotation and biases due to the difficulty of
measuring rotational broadening for low inclination objects, but these
are unlikely to change the results by more than 1-2 per cent. The
quoted uncertainties include the statistical precision, which is
dominant, and also contributions that account for any plausible
uncertainty in the cluster distance and age.

The distribution of projected radius in these low-mass Pleiades stars
is consistent with a random orientation of spin axes and is
inconsistent with strong alignments where the spin-axis vectors are
confined to cones with semi-opening angle $<30^{\circ}$. Weaker
alignments are possible if the mean inclination angle is $\leq
45^{\circ}$, but these scenarios would lead to a larger inferred over-radius.

There is no evidence that the radius inflation with respect to model
predictions varies over the luminosity range considered (approximately
$-0.7 \geq \log L/L_{\odot} \geq -2.7$) and in particular, no evidence
for a change for PMS stars that are fully convective. The same models
do predict radii that reasonably match the interferometrically measured
radii of older, magnetically inactive field stars with masses and
luminosities in the upper half of this range, which is circumstantial
evidence that magnetic activity or rapid rotation are the factors
responsible.  A comparison with existing ``magnetic models'' suggests
that neither magnetic inhibition of convection or flux blocking by
starspots can solely explain the over-radius at the expected levels of
surface magnetic field or spot coverage, however a simple combination
of the two effects does match the data quite well. One remaining puzzle
is that although all the stars we consider are very fast-rotating and
likely to have saturated levels of magnetic activity, there is evidence
that it is the slowest rotating half of this sample that have the
largest over-radii.

That low-mass, active stars have larger radii at a given luminosity
than predicted by the most commonly used evolutionary models has
several important implications. Effective temperatures would be lower;
ages derived using the Hertzsprung-Russell diagram and non-magnetic,
standard PMS isochrones would be underestimated, as would stellar
masses; core temperatures would be lower than expected, leading to
delays in the onset of lithium depletion and an extension of the PMS
lifetime. The calibration of these effects and the identification of
the causes of radius inflation requires careful observation and radius
measurements for stars at a range of masses in clusters covering the
full range of PMS evolution.

\section*{Acknowledgments}

Data presented herein were obtained at the WIYN 3.5m Observatory from
telescope time allocated to NN-EXPLORE through (a) the scientific
partnership of the National Aeronautics and Space Administration,
the National Science Foundation, and the National Optical Astronomy
Observatory, and (b) Indiana University's share of time on the WIYN
3.5-m. This work was supported by a NASA WIYN PI Data Award, administered by the
NASA Exoplanet Science Institute, though JPL RSA \# 1560105.
RJJ and RDJ also wish to thank the UK Science and Technology Facilities Council for financial support. 

\nocite{Rebull2016b}
\nocite{Morales2009a}
\nocite{Torres2013a}
\nocite{Feiden2014a}
\nocite{MacDonald2013a}
\nocite{Jackson2014a}
\nocite{Stauffer2003a}
\nocite{Covey2016a}
\nocite{Rebull2016a}
\nocite{Somers2014a}
\nocite{Jackson2009a}
\nocite{Jackson2016a}
\nocite{Hartman2010a}
\nocite{Lanzafame2017a}
\nocite{Baraffe2015a}
\nocite{Soderblom2009a}
\nocite{Stauffer1998a}
\nocite{Carpenter2001a}
\nocite{An2007a}
\nocite{Rieke1985a}
\nocite{Jackson2010a}
\nocite{Jackson2010b}
\nocite{Horne1986a}
\nocite{Bagnulo2003a}
\nocite{Claret1995a}
\nocite{Krishnamurthi1998a}
\nocite{Kenyon1995a}
\nocite{Skrutskie2006a}
\nocite{Melis2014a}
\nocite{vanLeeuwen2009a}
\nocite{Queloz1998a}
\nocite{Reinhold2013a}
\nocite{Stauffer1987a}
\nocite{ODell1994a}
\nocite{Marilli1997a}
\nocite{Feiden2012a}
\nocite{Feiden2015a}
\nocite{Boyajian2012a}
\nocite{Somers2015a}
\nocite{Somers2015b}
\nocite{Barnes2003a}
\nocite{Barnes2007a}
\nocite{Somers2015a}
\nocite{Dotter2008a}
\nocite{Fang2016a}
\nocite{Spruit1986a}
\nocite{Demory2009a}
\nocite{Malo2014a}
\nocite{Spruit1982a}
\nocite{vanSaders2012a}
\nocite{Kawaler1988a}
\nocite{Reiners2012a}
\nocite{Corsaro2017a}
\nocite{Jackson2010a}
\nocite{Douglas2017a}
\nocite{Howell2014a}
\nocite{Mermilliod1992a}
\nocite{Krause1980a}
\nocite{LopezMorales2005a}
\nocite{Feiden2016a}
\nocite{Messina2016a}
\nocite{Jeffries2017a}
\nocite{Somers2017a}
\nocite{Morales2008a}
\nocite{Mullan2001a}
\nocite{Kamai2014a}
\nocite{Kraus2015a}
\nocite{Kraus2016a}
\nocite{David2016a}
\nocite{Bershady2008a}
\nocite{Soderblom2009a}
\nocite{Soderblom1993a}
\nocite{Pecaut2013a}
\nocite{Jackson2014b}
\nocite{Jeffries2005a}
\nocite{Cummings2017a}
\nocite{Duchene2013a}
\nocite{Raghavan2010a}
\nocite{Rhode2001a}
\nocite{Jeffries2007b}
\nocite{Jackson2012a}
\nocite{Irwin2007a}

\bibliographystyle{mn2e} 
\bibliography{references}


\bsp 
\label{lastpage}
\end{document}